\documentclass[aip,amsmath,amssymb,reprint]{revtex4-2}
\usepackage{natbib}

\usepackage{amsmath, amssymb, amsthm}
\usepackage[all]{xy}
\usepackage{graphicx}
\usepackage{tikz}
\usetikzlibrary{matrix}
\usepackage{verbatim}
\usepackage{enumerate}
\usepackage{url}
\usepackage{subfigure}
\usepackage{bm}
\usepackage{times}
\usepackage{overpic}
\usepackage{tikz}
\usetikzlibrary{shapes.geometric,arrows}
\tikzstyle{arrow} = [thick,->,>=stealth]
\usepackage{hyperref}

\newcommand{\note}[1]{}

\usepackage{xcolor}

\renewcommand{\imath}{i}

\DeclareMathOperator{\codim}{codim}

\newcommand{\tr}{\textsf{T}}
\renewcommand{\u}{\mathbf{u}}

\newcommand{\Ab}{\mathbf{a}}
\newcommand{\Bb}{\mathbf{b}}

\newcommand{\Db}{\mathbf{d}}

\newcommand{\wb}{\mathbf{w}}
\newcommand{\vb}{\mathbf{v}}
\newcommand{\ub}{\mathbf{u}}

\renewcommand{\k}{w}
\newcommand{\kb}{\mathbf{w}}
\newcommand{\kbs}{\wb^\star}
\newcommand{\ks}{w^\star}

\newcommand{\e}{\varepsilon}
\newcommand{\eps}{\e}

\newcommand{\A}{\mathbf{a}}

\newcommand{\R}{\mathbb{R}}

\newcommand{\Z}{\mathbb{Z}}

\newcommand{\abs}[1]{\left|#1\right|}

\newcommand{\norm}[1]{\left\|#1\right\|}

\newcommand{\rset}[2]{\left\lbrace\, #1\,\left|\;#2\right.\right\rbrace}

\newcommand{\set}[2]{\rset{#1}{#2}}

\newcommand{\W}{\mathbb{W}}

\newcommand{\U}{\mathbf{U}}
\newcommand{\w}{\mathbf{w}}
\newcommand{\ws}{\w^\star}
\newcommand{\V}{\mathbf{V}}
\renewcommand{\v}{\mathbf{v}}

\DeclareMathOperator{\sgn}{sgn}

\newcommand{\Ct}{\mathbf{C}}

\begin{document}

\title
{Multiple Timescale Dynamics of Network Adaptation with Constraints}

\author{Erik A. Martens}
\affiliation{
Centre for Mathematical Sciences, Lund University, M\"arkesbacken 4, 223 62 Lund, Sweden\
}
\affiliation{Centre for Mathematical Modeling - Human Health and Disease, IMFUFA, Department of Science and Environment, Universitetsvej 1, Roskilde University, Roskilde, Denmark}
\author{Christian Bick}
\affiliation{
Department of Mathematics, Vrije Universiteit Amsterdam, De Boelelaan 1111, Amsterdam, The Netherlands
}
\affiliation{
Institute for Advanced Study, Technische Universit\"at M\"unchen, Lichtenbergstr 2, 85748 Garching, Germany
}
\affiliation{
Department of Mathematics, University of Exeter, Exeter EX4 4QF, United Kingdom
}
\affiliation{
Mathematical Institute, University of Oxford, Oxford OX2 6GG, United Kingdom
}

\date{\today}

\begin{abstract}
Adaptive network dynamical systems describe the co-evolution of dynamical quantities on the nodes as well as dynamics of the network connections themselves. 
For dense networks of many nodes, the resulting dynamics are typically high-dimensional.
Here we consider adaptive dynamical systems subject to constraints on network adaptation: Asymptotically, the adaptive dynamics of network connections evolve on a low-dimensional subset of possible connectivity. Such dimension reduction may be intrinsic to the adaptation rule or arise from an additional dynamical mechanism acting on a timescale distinct from that of network adaptation.
We illustrate how network adaptation with various constraints influences the dynamics of Kuramoto oscillator networks and elucidate the role of multiple timescales in shaping the dynamics.
Our results shed light on why one may expect effective low-dimensional adaptation dynamics in generally high-dimensional adaptive network dynamical systems.
\end{abstract}

\maketitle 
\allowdisplaybreaks

\begin{quotation}
Networks are pervasive in both nature and technology, encompassing systems of varying scales, from metabolic cell circuits to global transport systems~\cite{strogatz2001exploring,BarabasiOltvai2004}.
Dynamical processes involving networks lead to collective behavior, such as synchronization of oscillators~\cite{strogatz2004sync,PikvskyRosenblum2003synchronization}. 
Dynamics may occur on the nodes (dynamics \emph{on} the network) or on the edges themselves (dynamics \emph{of} the network). 
Co-evolutionary (or adaptive) networks exhibit dynamics both on and of the network, so that the structure of the network itself evolves in response to node states or external stimuli~\cite{gross2008adaptive,berner2023adaptive}. 
One of the key challenges in analyzing these systems is the sheer number of variables involved, especially in dense networks. 
To manage this complexity, recent efforts have focused on reducing the dimensionality, e.g., by considering the evolution of (row-)averaged weights in the connectivity matrix~\cite{Duchet2023,cestnik2025continuum}.
In this paper, we consider adaptive network dynamical systems whose edge adaptation dynamics are subject to constraints; this provides a mathematical framework to capture low-dimensional adaptation dynamics.
We analyze several examples of networks with constrained adaptation and compare their dynamics to unconstrained adaptation. 
Additionally, we investigate adaptation with constraints adhering to a non-zero time-scale separation. 
This systematic approach aims to elucidate the multiple timescale dynamics of adaptive networks and their implications for network functionality.
\end{quotation}

\section{Introduction}

\noindent
Networks are ubiquitous in nature and technology, spanning systems from small to large scales~\cite{strogatz2001exploring}. 
Examples include metabolic cell circuits, vascular networks, neural networks in the brain, power grids, and transport systems; cf.~Refs.~\onlinecite{BarabasiOltvai2004,kirst2020mapping,sporns2013structure,pagani2014power,Kaluza2010}.
On the one hand, one can consider dynamical processes where dynamical quantities evolve on the nodes (dynamics \emph{on} the network).
These give rise to fascinating collective phenomena, such as synchronization~\cite{strogatz2004sync,PikvskyRosenblum2003synchronization}; examples of dynamics on the network include neurons, heart cells, flashing fireflies, Josephson junctions, metronomes and pendulum clocks. 
On the other hand, the network connections themselves may be dynamic and change over time (dynamics \emph{of} the network).
Examples include changing friendships in social networks, synaptic plasticity in the brain, or regulation of flow in vascular and glymphatic networks~\cite{Kempter1999,SkyrmsPemantle2009,jacobsen2009tissue,mestre2020cerebrospinal}.

The co-evolution of dynamics on the network and of the network---or simply adaptive network dynamics---has attracted increased interest recently~\cite{gross2008adaptive,berner2023adaptive}.
The resulting dynamical systems are high dimensional and typically include various time scales, making them difficult to analyze mathematically.
Thus, research has focussed on using simple oscillator models~\cite{seliger2002plasticity,niyogi2009learning, aoki2011self,lucken2016noise,burylko2018winner,kim2021multifrequency,augustsson2024co}, such as the Kuramoto model, theta neuron model and others, or the dynamics in transport networks with adaptive edges subject to load fluctuations~\cite{martens2017transitions,klemm2023bifurcations}.
The curse of dimensionality provides a particular challenge: 
If the underlying network is dense, then an adaptive network with~$N$ nodes has~$\mathcal{O}(N^2)$ degrees of freedom.
To manage this complexity, recent efforts reduced the dimension by considering the dynamics of averages of adaptive edge weights.
For example, rather than having each connection weight~$w_{kj}$ evolve individually, one may take  inspiration from neuroscience and consider adaptive equations with a single input (or output) weight per node, to mimic homeostatic scaling~\cite{turrigiano2008self,maistrenko2014solitary}. The resulting system has~$\mathcal{O}(N)$ dynamical variables which reduces the complexity of the problem.

In this paper, we develop a systematic approach for adaptive network dynamical systems that are subject to constraints, which allow for an effective dimensionality reduction. 
More specifically, we focus on \emph{linear constraints}, i.e., a constraint that corresponds to an affine subspace~$\Ct\subset \R^{N\times N}$, see Sec.~\ref{sec:FormalEquations}. 
The dynamics of the connection weights can be described in terms of adaptation both along the constraint and transverse to it. Adaptation dynamics subject to constraints are then illustrated in Sec.~\ref{sec:TwoAdaptOsc} for small and large adaptive Kuramoto oscillator networks.
On the one hand, constraints may be \emph{intrinsic} to the adaptation dynamics, i.e., there is a constraint~$\Ct$ such that the transverse dynamics decays to zero:
The adaptation will (asymptotically) evolve on the low-dimensional constraint surface~$\Ct$ due to the intrinsic system properties.
On the other hand, constraints may be forced by a dynamical mechanism that acts on the transverse dynamics only---we refer to this as \emph{homeostasis}.
Introducing constraints this way yields slow-fast dynamics: 
Homeostasis that enforces the constraint evolves on a timescale that is fast compared to the adaptation. In the singular limit of infinitely fast adaptation, the dynamics evolve on the constraint~$\Ct$. In Sec.~\ref{sec:SubgraphConstr}, we consider networks of adaptive Kuramoto oscillators subject to subgraph and (weighted) degree constraints. We conclude with a discussion in Sec.~\ref{sec:Discussion}.

\section{Adaptive network dynamical systems with constraints\label{sec:constrained_adaptive_network_dynamics}}
\label{sec:FormalEquations}

\noindent
Consider a network dynamical system on~$N$ nodes, where the state $x_k\in\R^{d}$ of node $k\in\{1, \dotsc, N\}$ evolves according to
\begin{subequations}\label{eq:AdaptiveNetwork}
\begin{align}\label{eq:NodeDyn}
\dot x_k &= F_k(x_k) + \sum_{j=1}^N w_{kj}G(x_k,x_j),
\intertext{
with intrinsic node dynamics determined by a function~$F_k$ and the coupling determined by~$G$ (both sufficiently smooth functions) subject to weights~$w_{kj}$.
The network is \emph{adaptive} if the weight matrix $\w=(w_{kj})$ co-evolves with the node dynamics~\eqref{eq:NodeDyn} according to the adaptation equation}
\label{eq:AdaptFree}
\dot w_{kj} &= A(x_k,x_j,w_{kj})
\end{align}
\end{subequations}
determined by a sufficiently smooth adaptation function~$A$.

To simplify notation, we assume that $\dot w_{kk} = 0$, but that~$w_{kj}$ for $k\neq j$ is subject to adaptation.
In other words, adaptation describes the evolution on an $N(N-1)$-dimensional space of matrices
\begin{equation}
\W \subset \R^{N\times N}
\end{equation}
with fixed entries on the diagonal\footnote{One can assume $w_{kk}=0$ without loss of generality but other choices may be convenient depending on the system at hand.}.
The underlying network is \emph{fully adaptive} in the sense that any weight between distinct nodes is subject to adaptation.
It is straightforward to restrict to adaptation on arbitrary networks, where edges that are present are subject to adaptation.

\subsection{Example: Adaptive Kuramoto oscillator networks}

\noindent
A widely analyzed class of adaptive network dynamical systems are adaptive Kuramoto oscillator networks. 
Identifying the state variable~$x_k$ in Eqn.~\eqref{eq:AdaptiveNetwork} with the oscillation phase $\phi_k\in\mathbb{T}:=\R/2\pi\Z$ of node $k\in\{1,\ldots,N\}$, we obtain the adaptive Kuramoto oscillator model
\begin{subequations}\label{eq:full_system}
    \begin{align}
    \dot{\phi}_k &= \omega_k + \frac{1}{N}\sum_{j=1}^N w_{kj} \sin(\phi_j-\phi_k+\alpha) \label{eq:full_system_fi},\\
    \dot{w}_{kj} &= 
    \epsilon(b+a\cos(\phi_j-\phi_k+\beta)-w_{kj}) \label{eq:full_system_k},
    \end{align}
\end{subequations}
where $\omega_k$~is the intrinsic or natural frequency of oscillator~$k$, oscillators interact via sinusoidal interactions with phase-lag~$\alpha$, and adaptation acts on a time scale determined by $\epsilon>0$ with baseline (or standard coupling)~$b$, adaptation strength (or adaptivity)~$a$, and adaptation shift~$\beta$.
We set $w_{kk} = b+a\cos(\beta)$, the asymptotic value it would assume if the self coupling was adaptive.

This model is well-suited to illustrate the framework of constrained adaptation, as it has already been investigated using various assumptions including identical or distributed natural frequencies, zero phase-lag, and more~\cite{seliger2002plasticity,maistrenko2007multistability,berner2021desynchronization,berner2023adaptive,juttner2023complex,cestnik2025continuum,Duchet2023,sales2024recurrent,wei2024synchronization}. 

\subsection{Constraints and homeostasis}

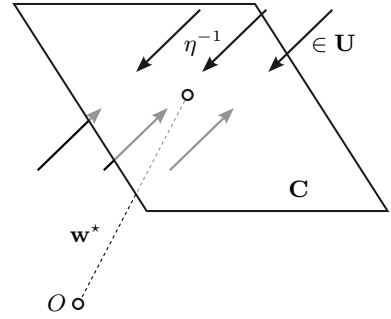
\begin{figure}
\centering
 \begin{overpic}[width=0.7\columnwidth]{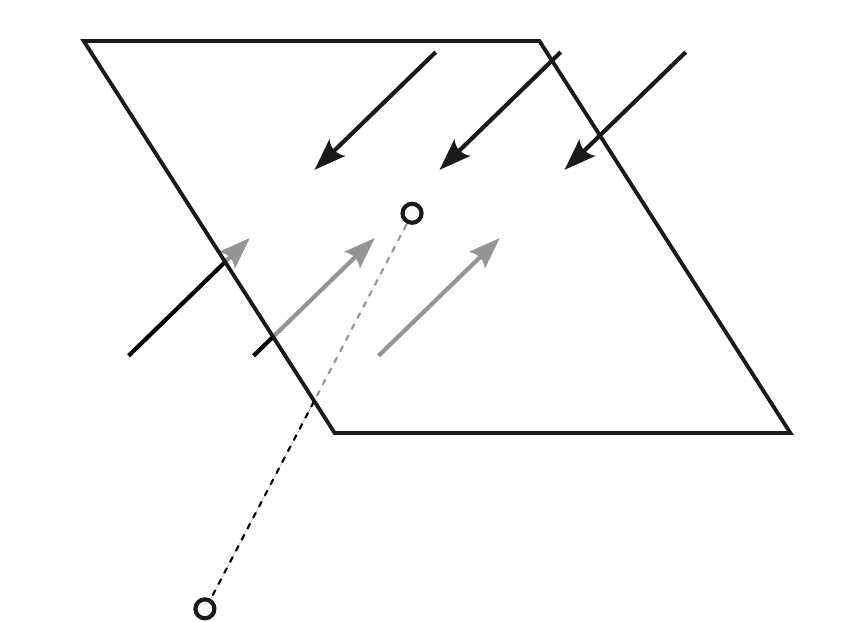}
  \put(17,0){$O$}
  \put(22,15){$\w^\star$}
  \put(47,57){$\eta^{-1}$}
  \put(75,57){$\in\U$}
  \put(70,25){$\Ct$}
 \end{overpic}
\caption{\label{fig:Constraint}
An $M$-dimensional constraint~$\Ct$ is an affine space offset from the ori gin~$O$ by~$\wb^\star$ with $M$-dimensional tangential space~$\V$ as illustrated in this schematic.
Homeostasis dynamics---indicated by arrows---acts transverse to~$\Ct$ along directions given by~$\U$ (i.e., the arrows as vectors are elements of~$\U$).
If this schematic was literally interpreted as a two-dimensional constraint in a three-dimensional space of weights then $\dim(\V)=2$ and $\dim(\U)=\codim(\V)=1$.
}
\end{figure}

\noindent
A (linear) \emph{$M$-dimensional constraint for the adaptation~\eqref{eq:AdaptFree}} is given by an affine subspace
\begin{equation}\label{eq:GenConstr}
\Ct = \V + \ws \subset \W,
\end{equation}
determined by $\ws\in\W$ and an $M$-dimensional linear subspace~$\V\subset\W$
tangent to~$\Ct$ with $M<N(N-1)=\dim(\W)$; see Fig.~\ref{fig:Constraint} for an illustration.
Regarding notation, $\v$~denotes an element (a weight matrix) of a linear subspace~$\V\subset\W$ of weight matrices---these can be identified with column vectors where convenient.

The space of matrices~$\W$ can now be split into a component along the constraint~$\Ct$ and transverse to it.
Specifically, write
\[\W = \V \oplus \U,\]
where~$\U$ a subspace of co-dimension~$N(N-1)-M$ transverse to the space~$\V$ tangent to the constraint~$\Ct$.
The choice of~$\U$ is arbitrary; we will typically consider~$\U=\V^\perp$, the orthogonal complement to~$\V$. 
Let~$\Pi_\V$ and~$\Pi_\U$ denote the projections onto~$\V$ (along~$\U$) and onto~$\U$ (along~$\V$), respectively.
Now we can decompose any weight matrix~$\w$ into
\begin{subequations}\label{eq:ConstrProj}
\begin{align}
\w_\V &:= \Pi_\V(\w-\ws)\\
\w_\U &:= \Pi_\U(\w-\ws),
\end{align}
\end{subequations}
which are the components parallel and transverse to the constraint~$\Ct$ relative to a translation by the constant~$\ws$.

We now consider the adaptation dynamics along and transverse to the constraint. 
Write---in slight abuse of notation---the adaptation~\eqref{eq:AdaptFree} as
\begin{equation}\label{eq:AdaptMatrix}
\dot \w = A(x;\w).
\end{equation}
We refer to the \emph{adaptation dynamics subject to the constraint~$\Ct$} as the system given by
\begin{subequations}\label{eq:AdaptConstr}
\begin{align}
\dot \w_\V &= \Pi_\V A(x,\w_\V,\w_\U)\label{eq:ConstrDyn}\\
\dot \w_\U &= \Pi_\U A(x,\w_\V,\w_\U)-\eta^{-1}\w_\U.\label{eq:Homeostasis}
\end{align} 
\end{subequations}
with \emph{constraint parameter}~$\eta \geq 0$.
In particular, the additional linear term in equation~\eqref{eq:Homeostasis} now corresponds to dynamics transverse to the constraint---we simply refer to this as \emph{homeostasis dynamics} which vanish on the constraint~$\Ct$.
Note that one could consider other functional forms (e.g., nonlinear) for the homeostasis.

The timescale separation parameter~$\eta$ yields distinct limits for the constraint.
First, in the limit $\eta\to \infty$ homeostasis vanishes and the system is equivalent to the original adaptive network dynamics~\eqref{eq:AdaptFree}.
Second, the limit $\eta\to 0$ corresponds to the homeostasis being infinitely fast so that~$\w_\U=0$. 
In this case, the adaptation dynamics reduce to dynamics on the constraint~$\Ct$ given by
\begin{align}\label{eq:AdaptConstrLim}
\dot \w_\V &= \Pi_\V A(x,\w_\V,0),
\end{align}
corresponding to adaptation dynamics with reduced dimension~$M<N(N-1)$.

One can in principle extend this approach to \emph{nonlinear constraints} that, for example, may be given by an algebraic curve. This is straightforward in a neighborhood of the constraint where directions tangential and transverse to the constraint manifold are well defined. 
Care has to be taken close to singular points of the constraint manifold. Moreover, the constraint dynamics in the limit $\eta\to 0$ may not globally reduce to the constraint; for an analog scenario see Ref.~\onlinecite{Guckenheimer1975}.

\subsection{Example constraints\label{sec:constraint_examples}}

\noindent
To illustrate the versatility of constraints, we give examples of constraints for the adaptive dynamics and their relationship to previous research. Recall that we exclude adaptation of self-links, so there are $N(N-1)$ real-valued dynamical weights.

First, there are \emph{network-level constraints}.
Adaptive dynamics with \emph{global adaptive coupling with adaptive weight~$w\in\R$} corresponds to a constraint
\begin{equation}\label{eq:constraint_global_adaptive}
\Ct = \set{\w=(w_{kj})}{w_{kj}=w}.
\end{equation}
The resulting constrained adaptation dynamics are one-dimensional (i.e., $M=1$).
Adaptive network dynamical systems with a single adaptive weight have been considered in Ref.~\onlinecite{Duchet2023}. 
This setup can be easily generalized to coupled populations with non-uniform coupling (e.g, chimera states were studied in setups where the coupling weights within each population are the same, while the coupling between populations are the same, but distinct from the coupling within each population~\cite{Bick2018c}.).

Another network-level constraint is \emph{adaptation subject to a constant mean coupling weight~$c$} corresponds to a constraint
\begin{equation}
\Ct = \set{\w=(w_{kj})}{\textstyle\sum_{j\neq k}w_{kj}=c}
\end{equation}
of {co-dimension 1 (i.e., $M=N(N-1)-1$).} 
This constraint can be motivated by neural dynamics, where a homeostasis mechanism keeps the overall network coupling weight constant; cf.~Refs.~\onlinecite{turrigiano2008self,goriely2020neuronal}.
While this constraint considers the average of the coupling weights, it is possible to generalize this constraint to involve general (linear) functions of the coupling weights.

Second, in analog to the network-level constraints there are natural \emph{node-level constraints}. 
Adaptive dynamics with \emph{a single adaptive input/output weight $w_k\in\R$ per node} correspond to the constraints
\begin{subequations}
\begin{align}
\Ct &= \set{\w=(w_{kj})}{w_{kj}=w_k}\label{eq:constraint_adaptive_input},\\
\Ct &= \set{\w=(w_{kj})}{w_{kj}=w_j}\label{eq:constraint_adaptive_output},
\end{align}
\end{subequations}
respectively.
The resulting constrained weight matrices have constant row and column elements, respectively, and so that the adaptive dynamics evolve on a constraint of dimension~$M=N$. 
These constraints yield adaptive versions of the network dynamics considered in Refs.~\onlinecite{Hong2011a,maistrenko2014solitary}.
The resulting dynamics of the row average on the constraint manifold is subject of a recent study~\cite{cestnik2025continuum} and related work~\cite{carballosa2023cluster}.

Other relevant node-level constraints are \emph{constant input/output weights~$c_k$} per node, which corresponds to the constraints
\begin{subequations}\label{eq:RowSumConstr}
\begin{align}
\Ct &= \set{\w=(w_{kj})}{\textstyle\sum_{j\neq k}w_{kj}=c_k},\\
\Ct &= \set{\w=(w_{kj})}{\textstyle\sum_{j\neq k}w_{kj}=c_j},
\end{align}
\end{subequations}
respectively. 
The resulting constrained weight matrices have constant row and column sum, respectively, and are of {co-dimension~$N$ (i.e., $M=N(N-2)$).}
Again, this can be interpreted as a homeostasis mechanism that balances the input/output of each node to keep it constant.

Third, one may view \emph{adaptation on a subgraph} as a constraint.
Identify a given graph by its adjacency matrix $\A=(a_{kj})$.
The \emph{subgraph constraint corresponding to~$\A$} is given by
\begin{equation}
\Ct = \set{\w=(w_{kj})}{w_{kj}=0 \text{\ if\ }a_{kj}=0}.
\end{equation}
Here, the homeostasis mechanism lets the weights corresponding to edges that are not in~$\A$ decay to zero.
{The resulting constraint dynamics has dimension~$M$ equal to the number of edges in~$\A$.}

\section{Constraints for two adaptive Kuramoto oscillators}
\label{sec:TwoAdaptOsc}

\noindent
To illustrate constrained adaptation, let us consider the adaptive Kuramoto model Eq.~\eqref{eq:full_system} with $N=2$ oscillators. To simplify notation, write $\k_1:=w_{12}$, $\k_2:=w_{21}$ and identify
\[\wb = \left(
\begin{array}{cc}0 & w_{12}\\ w_{21} & 0\end{array}
\right)
\equiv
\left(\begin{array}{c}\k_1 \\ \k_2\end{array}\right)\]
with a column vector.
Introducing the phase difference $\phi:=\phi_1-\phi_2$, the adaptive network dynamics evolve according to
\begin{subequations}\label{eq:N2adaptiveKM}
\begin{align}
\dot{\phi} &= \omega + \frac{1}{2}\k_1\sin(\alpha-\phi) - \frac{1}{2}\k_2\sin(\alpha+\phi),\label{eq:governingRescalPhi}\\
\dot \k_1 &= \epsilon(b + a \cos(\beta+\phi)-\k_1),\label{eq:governingRescalKap12}\\
\dot \k_2 &= \epsilon(b + a \cos(\beta-\phi)-\k_2).\label{eq:governingRescalKap21}
\end{align}
\end{subequations}
where~\eqref{eq:governingRescalKap12} and~\eqref{eq:governingRescalKap21} give~\eqref{eq:AdaptMatrix}.
The full dynamics of this system was considered in Ref.~\onlinecite{juttner2023complex}. While the parameter~$\epsilon$ defines a second time-scale separation in addition to~$\eta$, we fix it at $\epsilon=0.1$ but retain it for comparison with existing results.

\subsection{Adaptive Kuramoto oscillators subject to a constraint}

\noindent
A general constraint~$\Ct$ as given in~\eqref{eq:GenConstr} for the system given by Eqs.~\eqref{eq:N2adaptiveKM} is determined by a direction~$\v = (v_1, v_2)^\tr\neq 0$ and an offset~$\kbs=(\ks_1, \ks_2)^\tr$.
Write $\langle\ ,\ \rangle$ for the usual Euclidean inner product and $\norm{\v}^2 = \langle\v,\v\rangle$.
A general transverse direction~$\U$ off the constraint space is given by $\u = (u_1, u_2)^\tr$.
Without loss of generality, we may assume $\norm{\vb}=\norm{\ub}=1$ and $\langle\vb,\kbs\rangle = 0$ (i.e., $\kbs$~is perpendicular to~$\Ct$). 
Write $\langle \ub,\kbs\rangle = \k$.
The projected variables~\eqref{eq:ConstrProj} become
\begin{subequations}\label{eq:ConstrProjPh0}
\begin{align}
\k_\V &= \langle\v, \kb-\kbs \rangle = v_1\k_1+v_2\k_2\\
\k_\U &= \langle\u, \kb-\kbs \rangle = u_1\k_1+u_2\k_2-\k.
\end{align}
\end{subequations}
(Since $\w_\V, \w_\U$ in~\eqref{eq:ConstrProj} are scalar quantities here, which can be seen as components of a transformed two-dimensional vector, we write them as $\k_\V, \k_\U$.)
Note that Eqs.~\eqref{eq:N2adaptiveKM} are written in terms of $\k_1$ and $\k_2$, while constraints are written in terms of~$\k_\V$ and~$\k_\U$.

Solving \eqref{eq:ConstrProjPh0} for $\k_1, \k_2$,
yields the inverted affine coordinates,
\begin{subequations}\label{eq:Inversion}
\begin{align}
    \k_1 &= 
    \frac{1}{D(\ub)}(- u_2 \k_\V + v_2 \k_\U  + v_2 \k),\\
    \k_2 &= 
    \frac{1}{D(\ub)}(u_1 \k_\V  - v_1 \k_\U  - v_1 \k)
\end{align}
\end{subequations}
with $D(\ub)=  u_2v_1- u_1v_2$. 
This allows us to rewrite the dynamical equations~\eqref{eq:N2adaptiveKM} in terms of~$\k_\V$ and~$\k_\U$:
Differentiating~\eqref{eq:ConstrProjPh0} yields
\begin{subequations}
\begin{align} 
    \dot\k_\V &= v_1\dot\k_1+v_2\dot\k_2\\
    \dot\k_\U &= u_1\dot\k_1+ u_2\dot\k_2.
\end{align}
\end{subequations}
Using~\eqref{eq:Inversion} and~\eqref{eq:N2adaptiveKM} to eliminate all dependencies on~$\k_1$ and~$\k_2$ we obtain the unconstrained dynamics
\begin{subequations}\label{eq:adaptiveKMN2_transformed}
\begin{align}
    \begin{split}
    \label{eq:kappaV}
    \dot\k_\V &=  H_\V(\k_\V,\k_\U) \\
    & = \epsilon((v_1+v_2)b + aP\cos(\phi+Q)-\k_\V)\
    \end{split}
    \\
    \begin{split}
    \label{eq:kappaU}
    \dot\k_\U &=  H_\U(\k_\V,\k_\U) \\
    &= \epsilon((u_1+u_2)b + aR\cos(\phi+S)-(\k_\U+\k)).
    \end{split}
    \end{align}
\end{subequations}
with $P=\sgn(P')\sqrt{P'^2+Q'^2}$, $Q=\arctan(-Q'/P')$ for $P' = (v_1+v_2)\cos(\beta)$, $Q'=(-v_1+v_2)\sin(\beta)$.
and $R = \sgn(R')\sqrt{R'^2+S'^2}$, $S=\arctan(-S'/R')$ for $R' = (u_2+u_1)\cos(\beta)$, $S'=-(u_1-u_2)\sin(\beta)$.
This describes the adaptive network dynamics without homeostasis, $\eta=\infty$.

Together, the general adaptation dynamics subject to the constraint~$\Ct$ for general~$\eta$ are now given by
\begin{subequations}\label{eq:ConstrDynKuram}
\begin{align}
\begin{split}
    \label{eq:phi_transformed}
    \dot{\phi} &= \omega + \frac{1}{2D(\ub)}(-u_2\k_\V+v_2 \k_\U + v_2\k)\sin(\alpha-\phi)
    \\&\qquad\quad -\frac{1}{2D(\ub)}(u_1 \k_\V - v_1 \k_\U-v_1\k)\sin(\alpha+\phi),
\end{split}
\\
\dot\k_\V &= H_\V(\k_\V,\k_\U),\label{eq:KuramConstrLim}\\
\dot\k_\U &= H_\U(\k_\V,\k_\U)-\eta^{-1}\k_\U.
\end{align}
\end{subequations}
Infinitely fast homeostasis with $\eta\to 0$ implies that $\k_\U\to 0$.
The adaptation of the coupling is in this limit given by Eq.~\eqref{eq:phi_transformed} and \eqref{eq:KuramConstrLim} with $\k_\U=0$ (see also Eq.~\eqref{eq:AdaptConstrLim}).
We discuss the resulting system in Sec.~\ref{sec:KMN2_forcedconstraints}.

\subsection{Intrinsic constraints for $N=2$ oscillators}\label{sec:N2_intrinsic_constraints}

\noindent
First, we assume that there is no homeostasis ($\eta=\infty$) and look for a constraint~$\Ct$ that is \emph{intrinsic} to~\eqref{eq:N2adaptiveKM}.
More precisely, we see an arbitrary constraint~$\Ct$ as a parameter and identify parameter values for which the dynamics asymptotically converge to~$\Ct$.
This generalizes results in Ref.~\onlinecite{juttner2023complex} to the setup of adaptive network dynamical systems with constraints.

Consider the dynamics along the direction orthogonal to~$\v$, that is, 
\begin{align*}
    \u = (u_1, u_2)^\tr = (-v_2, v_1)^\tr. 
\end{align*}
With $k = \langle \u, \kbs\rangle$, the constraint is given by
\begin{equation}
\Ct = \{\langle \u, \kb \rangle=\langle \u, \kbs \rangle=\k\},
\end{equation}
a line in~$\R^2$ through~$\kbs$ in the direction of~$\v$ determined by the normal vector~$\u$.

\begin{figure*}
\centering
    \begin{overpic}[width=0.32\textwidth]{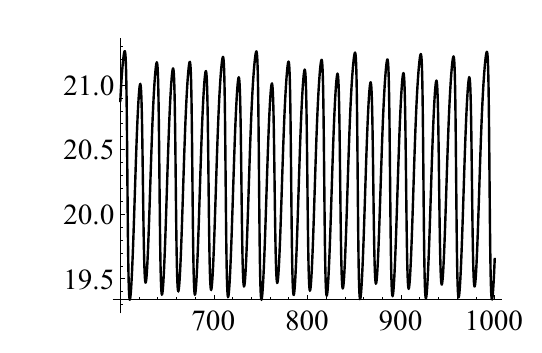}
        \put(10,60){(a1)}
        \put(100,7){$t$}
        \put(20,60){$\phi$}
        \put(50,60){\footnotesize $\eta=10^5$}
        \put(40,-5){\footnotesize Chaotic libration}
    \end{overpic}
    \begin{overpic}[width=0.32\textwidth]{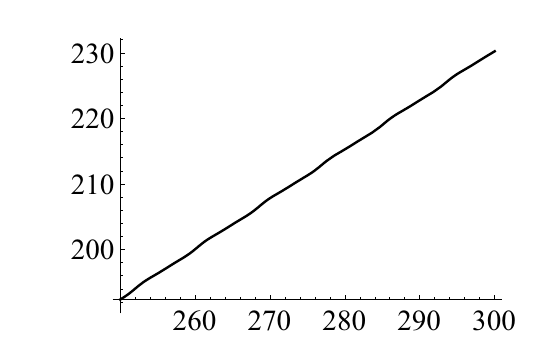}
        \put(10,60){(b1)}
        \put(100,7){$t$}
        \put(20,60){$\phi$}
        \put(50,60){\footnotesize$\eta=0$}
        \put(40,-5){\footnotesize Period-1 Libration}
    \end{overpic}
    \begin{overpic}[width=0.32\textwidth]{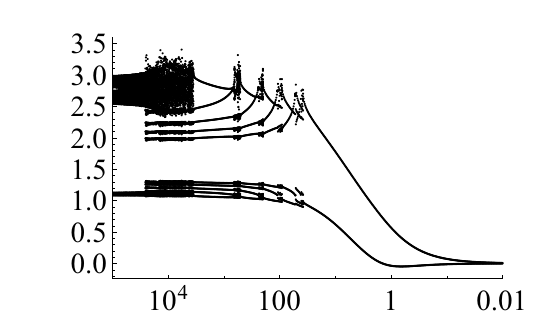}
        \put(10,60){(c1)}
        \put(100,5){$\eta$}
        \put(15,55){min$_t(\k_\U$), max$_t(\k_\U)$}
    \end{overpic}
    \\ 
    \begin{overpic}[width=0.32\textwidth]{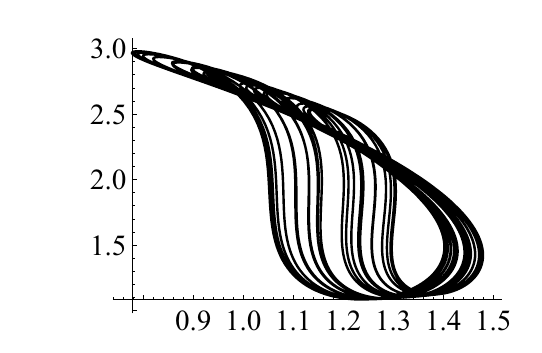}
        \put(10,60){(a2)}
        \put(100,10){$\k_\U$}
        \put(20,60){$\k_\V$}
    \end{overpic}
    \begin{overpic}[width=0.32\textwidth]{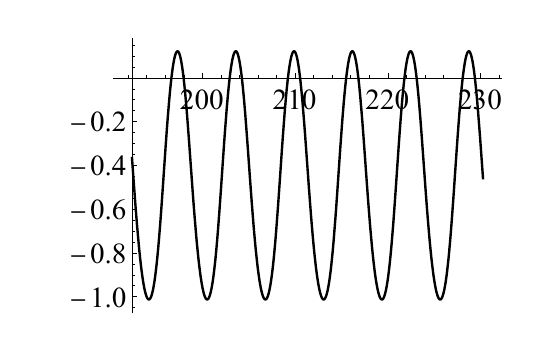}
        \put(10,60){(b2)}
        \put(100,50){$\phi$}
        \put(20,60){$\k_\V$}
    \end{overpic}
    \begin{overpic}[width=0.32\textwidth]{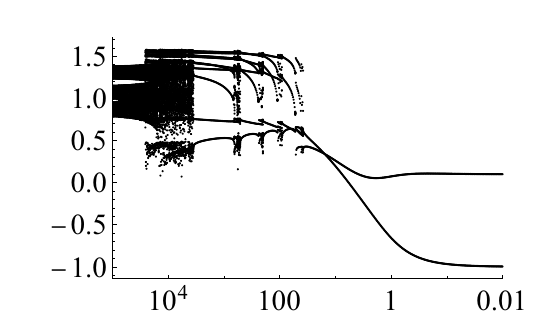}
        \put(10,60){(c2)}
        \put(100,5){$\eta$}
        \put(15,55){$\k_\V$}
    \end{overpic}
 \caption{\label{fig:squeezingchaos}
 The constraint enforced by homeostasis with time scale parameter~$\eta^{-1}$ effectively reduces the dimensionality of the dynamics of the system~\eqref{eq:ConstrDynKuram}.
 Column (a): Chaotic librations are observed for $\eta=10^5$, corresponding to the (effectively) unconstrained system (see also Ref.~\onlinecite{juttner2023complex}, Fig.~6(d)) for $\phi(t)$ (panel~(a1) and for corresponding values $(\k_\V(t),\k_\V(t))$ (panel (a2)).
 Column (b): Period-1 librations are observed for the hard constraint with $\eta = 0$ (see \eqref{eq:ConstrDynKuram} and following), for $\phi(t)$ (panel (b1) and for corresponding values $(\k_\V(t),\phi_V(t))$ (panel (b2)).
 Column (c): Decreasing $\eta $ from left to right reveals a transition from chaotic to period-1 motion.
 This sweep uses fixed initial conditions and reports minima/maxima in time of $\k_\U$ (panel (c1)) and corresponding values of $\k_\V$ at the same times (panel (c2)).
 Eqs.~\eqref{eq:ConstrDynKuram} are integrated with non-intrinsic constraint parameters $u_1=1,u_2=1/2,\k=0$ and system parameters
 $\alpha = -\pi/10,\beta = \pi /4,a =  -1.9425,\epsilon =0.2,b=1$, respectively.
 For further details and numerical procedures, see Sec.~\ref{sec:KMN2_forcedconstraints} .
 }
\end{figure*}
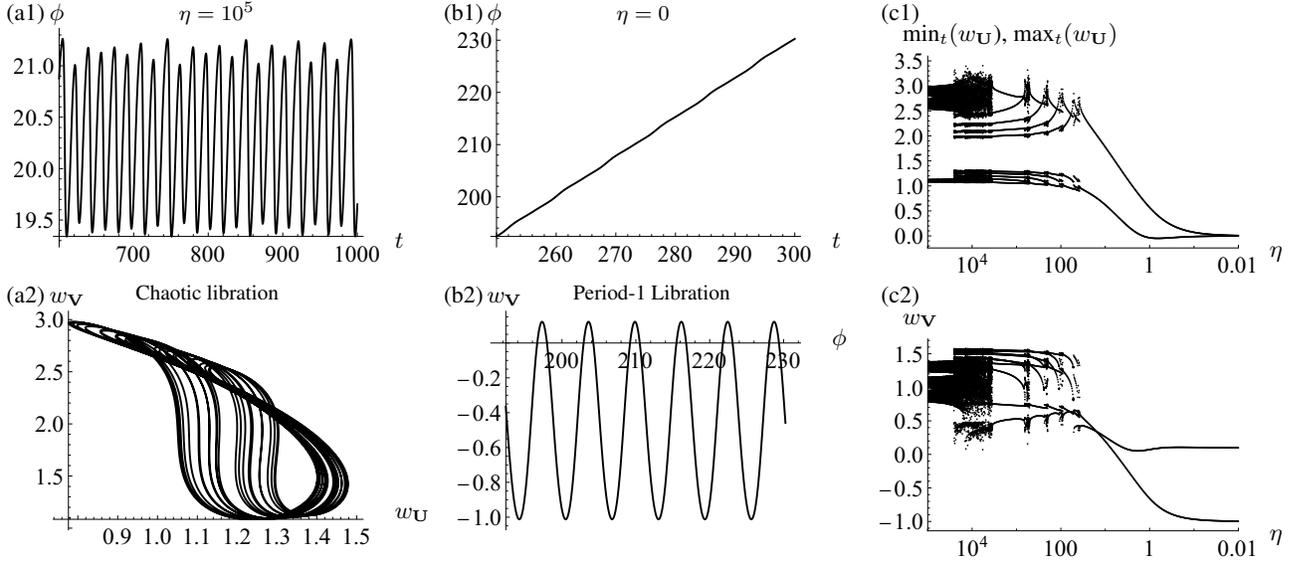

Decoupling of dynamical equations is a way in which how intrinsic constraints can arise.
First, note that the adaptation dynamics of~$\k_\U$ transverse to~$\Ct$ and of~$\k_\V$ along~$\Ct$ are not directly coupled; the only coupling is indirectly through the phase difference~$\phi$.
A sufficient condition for the transverse dynamics to decouple from the phase dynamics is given by
\begin{align}\label{eq:KMN2_intrinsic_constraint_dynamics_condition}
    0 &= \abs{R}^2 = 1-2u_1u_2(1-2\cos^2(\beta)).
\end{align}
If this condition is met, the transverse dynamics converges to a linearly stable equilibrium, given by
\[\ks_\U = b(u_1+u_2) - \k.\]
This means that the transverse dynamics~\eqref{eq:kappaU} in the direction $\ub = (u_1,u_2)$ vanish for any point in 
    \[\Ct = \V + \ks_\U\ub,\] 
since for $\kb\in\Ct$, we have $\Pi_\U(\kb) = \ks_\U$.
Thus, the set~$\Ct$ is an intrinsic constraint for parameters satisfying \eqref{eq:KMN2_intrinsic_constraint_dynamics_condition}.

The transverse dynamics~\eqref{eq:kappaU} now allow us to identify intrinsic constraints in the absence of homeostasis, $\eta=\infty$.
Since $\norm{\u}=1$ we have $|u_1|=|u_2|=1/\sqrt{2}$, and we see that the condition~\eqref{eq:KMN2_intrinsic_constraint_dynamics_condition} is satisfied for the following two cases, which align with results studied previously~\cite{juttner2023complex}:

\paragraph{Symmetric coupling: $u_1=-u_2$ and $\beta=m\pi$, $m\in\Z$:}
In this case, the transverse adaptation dynamics~\eqref{eq:kappaU} has a linearly stable equilibrium
\[\ks_\U =  - \k.\]
With the transformation~\eqref{eq:ConstrProjPh0},
we have $\k_\U = 1/\sqrt{2}(\k_1-\k_2)-\k$, we see that the adaptation relaxes on the diagonal in~$\W$ where weights are identical, $\k_1=\k_2$, i.e., the coupling difference converges to zero.

\paragraph{Offset antisymmetric coupling: $u_1=u_2$ and $\beta=\pi/2 + m\pi$, $m\in\Z$:}
In this case, the transverse adaptation dynamics~\eqref{eq:kappaU} has a linearly stable equilibrium
\[\ks_\U =  \sqrt{2}b-\k.\]
With the transformation~\eqref{eq:ConstrProjPh0},
we have $\k_\U = 1/\sqrt{2}(\k_1+\k_2)-\k$, we find that the adaptation relaxes to the weight configuration in~$\W$ with $\k_1+\k_2=2b$, i.e., the coupling sum converges to a constant.
In other words, the coupling strengths are antisymmetric with an offset~$2b$.

One may of course also consider an arbitrary transverse direction~$\u$ for a constraint.
Then the projections depend on~$\vb, \ub$ and the corresponding dynamics will generally contain both~$\k_\V, \k_\U$.
The normal transverse direction leads to a decoupling of the equations and yield a natural coordinate system.

\subsection{Forced constraints}
\label{sec:KMN2_forcedconstraints}

\noindent
While constraints may be intrinsic to the system, nonzero homeostasis $\eta\geq 0$ in~\eqref{eq:AdaptConstr} will enforce an arbitrary constraint~$\Ct$ in the limit $\eta\to 0$.
To illustrate the effect of homeostasis for a forced constrained we consider the network of two adaptive Kuramoto oscillators~\eqref{eq:N2adaptiveKM}.
Specifically, we focus on how adaptation constraints quench the chaotic librations for parameters $\alpha = -\pi/10,\beta = \pi /4,a =  -1.9425,\epsilon =0.2,b= 1$, and $\omega = 0.6944$ that were previously reported in Ref.~\onlinecite{juttner2023complex} (see Fig.~6 (d) therein).
First, we note that (effectively) unconstrained dynamics of Eqs.~\eqref{eq:ConstrDynKuram} occurs with a constraint parameter set to $\eta=10^5$, as shown in Fig.~\ref{fig:squeezingchaos}(a).

Next, we enforce a constraint determined by~$\vb$, normal transverse direction~$\ub$ and~$\k$, where we choose the constraint parameters $u_1=1, u_2=1/2$ and $k=0$. In the limit $\eta=0$, the constrained dynamics are given by Eq.~\eqref{eq:phi_transformed} and \eqref{eq:KuramConstrLim} with $\k_\U=0$, that is,
\begin{subequations}\label{eq:N2ForcedConstr}
\begin{align}
\begin{split}
\dot{\phi} &= \omega + \frac{1}{2D(\ub)}(-u_2\k_\V+u_1\k)\sin(\alpha-\phi)
\\&\qquad\quad -\frac{1}{2D(\ub)}(u_1\k_\V-u_2\k)\sin(\alpha+\phi),
\end{split}
\label{eq:PhaseForcedConstr}\\
  \dot{\k}_\V &=\epsilon((u_2-u_1)b + aP\cos(\phi+Q))-\k_\V.
\end{align}
\end{subequations}
These equations describe the dynamics on the constraint~$\Ct$, which are strictly two-dimensional and therefore cannot produce any chaotic dynamics.
The resulting dynamics for this set of parameters is oscillatory with period-1, as shown in Fig.~\ref{fig:squeezingchaos}(b).
The constrained adaptation effectively reduces the dimensionality of the attractors possible.

Bifurcations happen for finite $\eta>0$, as shown in Fig.~\ref{fig:squeezingchaos}(c).
To produce this bifurcation diagram, we gradually decreased~$\eta$ on a logarithmic scale, i.e., $\eta = 10^{p}$ where $p \in[-2,5]$ with even spacing of 0.01, and using identical initial conditions $(\phi,\k_\V,\k_\U)=(\pi,1.13,2.56)$.
For each value of~$\eta$, we report local minima/maxima in time for~$\k_\U(t)$ (panel (c1)), and corresponding values of~$\k_\V(t)$ (panel (c2)), after a transient time of 500 time units has passed.
As $\eta$ decreases, one observes a transition through various dynamic behaviors, starting from a chaotic region followed by a period doubling cascade and a region with mixed time scale phase slip dynamics. For approximately $\eta<12$, complicated oscillations cease to exist and instead a period-1 oscillation is observed (panels~(b)), congruent with the strongly constrained dynamics obtained as $\eta\to 0$.

\section{Constraints for higher-dimensional adaptive Kuramoto model}
\label{sec:SubgraphConstr}

\subsection{Subgraph constraints\label{sec:subgraph_constraints}}

\noindent
Recall that a subgraph constraint is determined by an adjacency matrix~$\A=(a_{kj})$ and given by
\begin{equation*}
    \Ct = \set{\w=(w_{kj})}{w_{kj}=0 \text{\ if\ }a_{kj}=0}.
\end{equation*}
The adaptive network dynamics subject to the constraint is given by
\begin{subequations}\label{eq:constrainedAdaptiveKM}
    \begin{align}\label{eq:constrainedAdaptiveKM_phi}
    \dot{\phi}_k &= \omega_k + \frac{1}{N}\sum_{j=1}^N w_{kj} \sin(\phi_j-\phi_k+\alpha), \,\\
    \label{eq:constrainedAdaptiveKM_w}
    \dot{w}_{kj} &= A(\phi_k,\phi_j, w_{kj}) - \eta^{-1}(1-a_{kj})w_{kj},  
    \end{align}
\end{subequations}
with adaptation function~$A$ given by~\eqref{eq:full_system_k}.
Enforcing the constraint ($\eta \to 0$) allows only the weights with $a_{kj}=1$ to be adaptive; the dynamics of all other edges $(k',j')$ is suppressed and $w_{k'j'}(t)\to 0$ as $t\to\infty$.  When the constraint is absent ($\eta\to\infty$), all edges adapt and co-evolve with the phases. Note that it is also possible to generalize the constraint by replacing the adjacency matrix $a_{kj}$ with a weighted matrix encoding weighted subgraph structure.

We illustrate the multiple timescale dynamics for networks of $N=2Q$ adaptive Kuramoto oscillators and a constraint that splits the network in two (disconnected) populations of $Q$~oscillators each.
More specifically, we consider the subgraph constraint determined by the adjacency matrix
\begin{align*}
        \A = 
        \left(
        \begin{array}{cc}
         \boldsymbol{1} & \boldsymbol{0}\\
         \boldsymbol{0} & \boldsymbol{1}
        \end{array}
        \right) \in \R^{N\times N}, 
\end{align*}
where $\boldsymbol{0}, \boldsymbol{1}$ are $Q\times Q$ matrices with coefficients $\boldsymbol{1}_{kj}=1$ and $\boldsymbol{0}_{kj}=0$.
We first analyze the extreme cases of no constraint ($\eta=\infty$) and the limit of an hard/enforced constraint ($\eta\to 0$) before considering finite time scale separation.
Note that while varying~$\eta$ the effective coupling strength between oscillators changes:
In the absence of a constraint we have a single population of $N$~globally coupled Kuramoto oscillators with coupling normalized by~$1/N$.
With a hard constraint we have two uncoupled populations of $Q$~globally coupled adaptive Kuramoto oscillators, each with coupling normalized by~$1/N=\frac{1}{2}(1/Q)$.
Thus, the uncoupled populations of $Q$~oscillators experience half the coupling strength relative to the size~$Q$ of the population.
Consequently, increasing~$\eta$ one may expect a transition from phase incoherence to phase coherence as the effective coupling is increased.

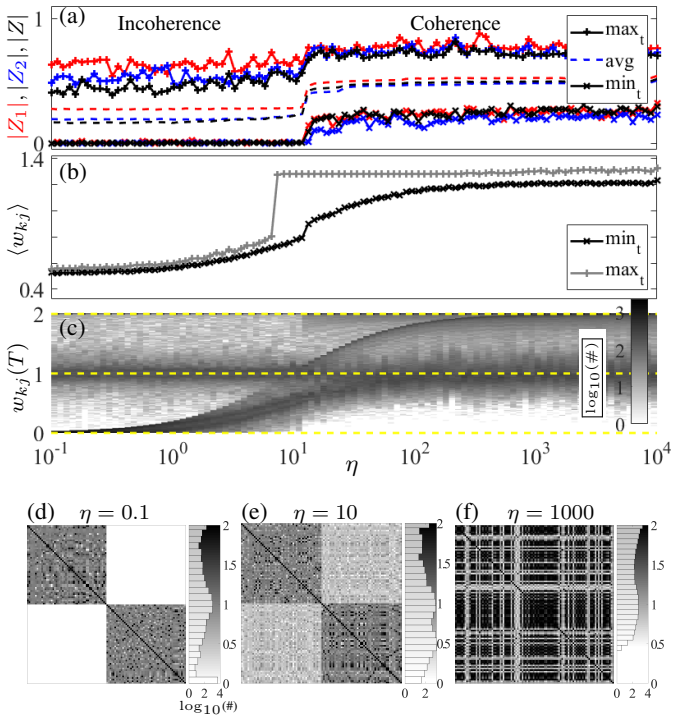
\begin{figure}
    \begin{overpic}[width=\columnwidth]{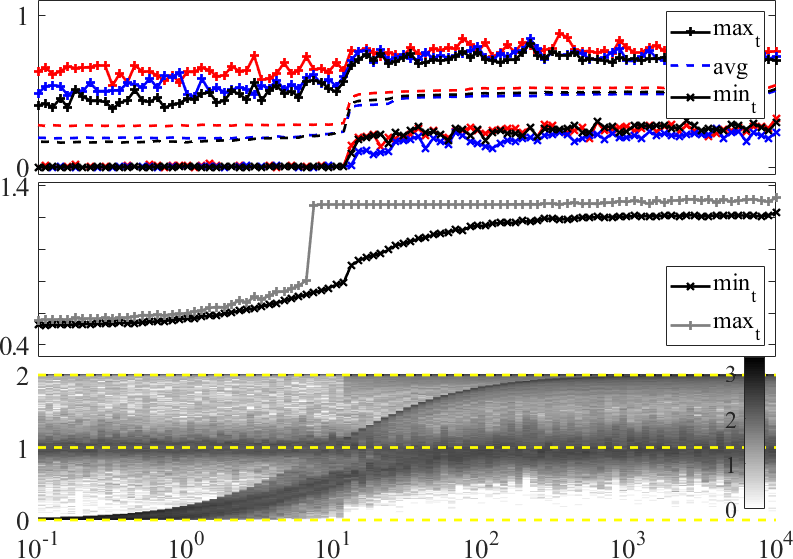}
    \put(50,-1){$\eta$}
    \put(15,67){\footnotesize Incoherence}
    \put(60,67){\footnotesize Coherence}
    \put(-2,50){\rotatebox{90}{\footnotesize $\textcolor{red}{|Z_{1}|},|\textcolor{blue}{Z_2}|, |Z| $}}
    \put(-2,32){\rotatebox{90}{\footnotesize $\langle w_{kj}\rangle$}}
    \put(-2,8){\rotatebox{90}{\footnotesize $w_{kj}(T)$}}
    \put(6,68){(a)}
    \put(6,44){(b)}
    \put(6,20){(c)}
    \setlength{\fboxsep}{1pt}
    \put(85,6){ \rotatebox{90}{ \colorbox{white}{\fbox{\textcolor{black}{\tiny $\log_{10}(\#)$}} }}}
    \end{overpic}
    \vspace{1em}
    \\
    \begin{overpic}[width=0.32\columnwidth]{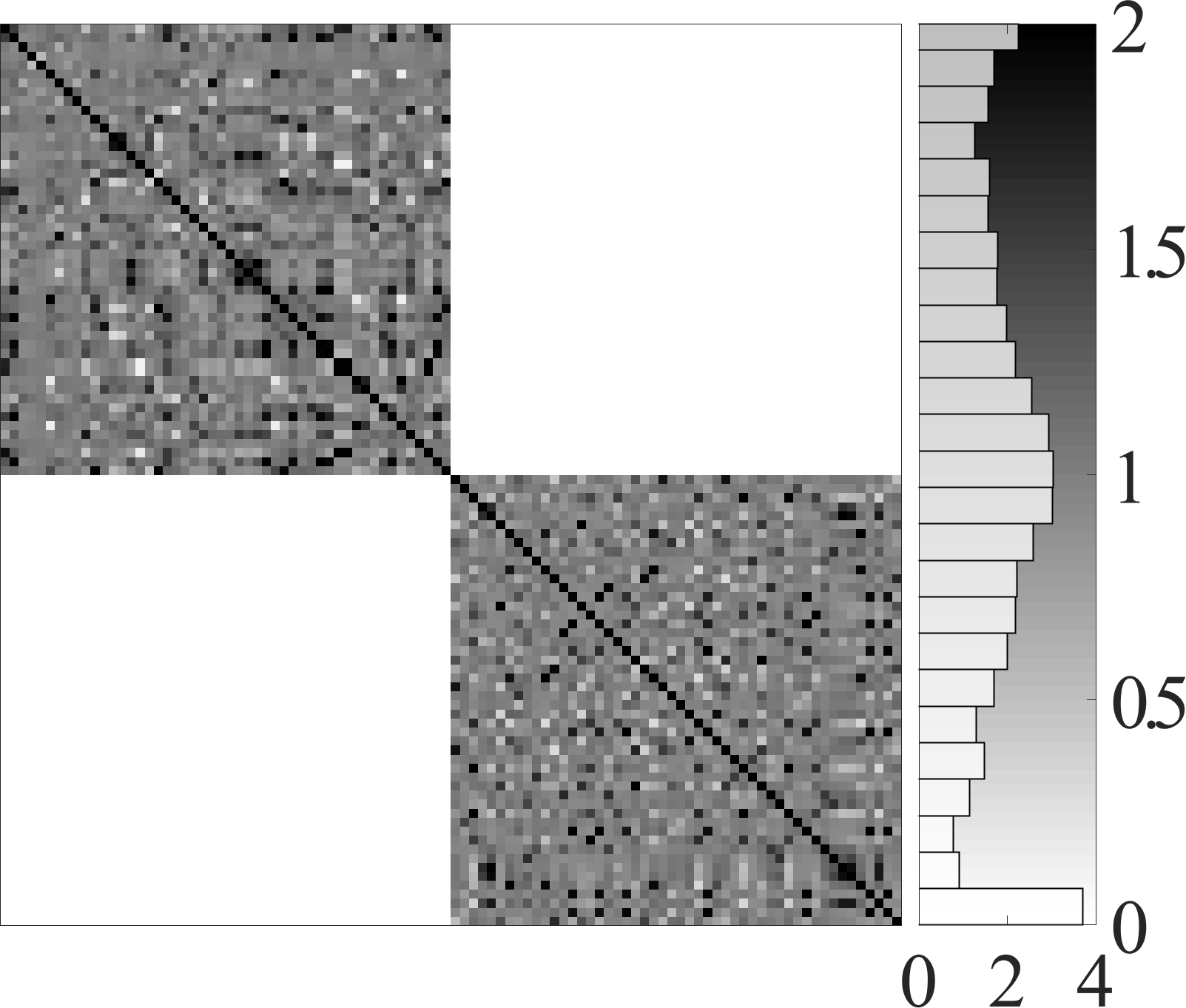}
        \put(25,86){\footnotesize $\eta = 0.1$}
        \put(0,86){(d)}
        \put(73,-6){\tiny $\log_{10}$(\#)}
    \end{overpic}
    \begin{overpic}[width=0.32\columnwidth]{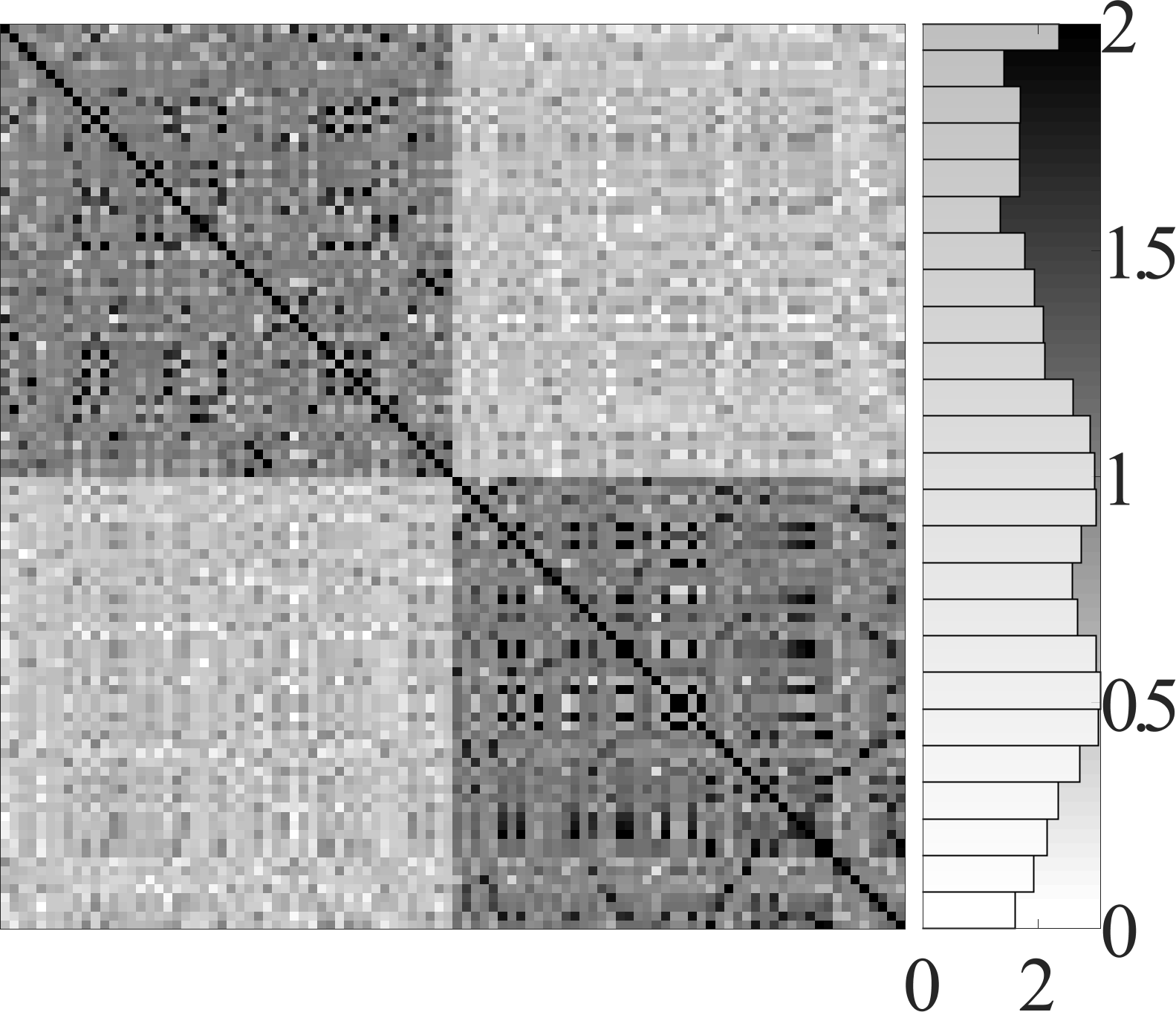}
        \put(25,86){\footnotesize $\eta = 10$}
        \put(0,86){(e)}
    \end{overpic}
    \begin{overpic}[width=0.32\columnwidth]{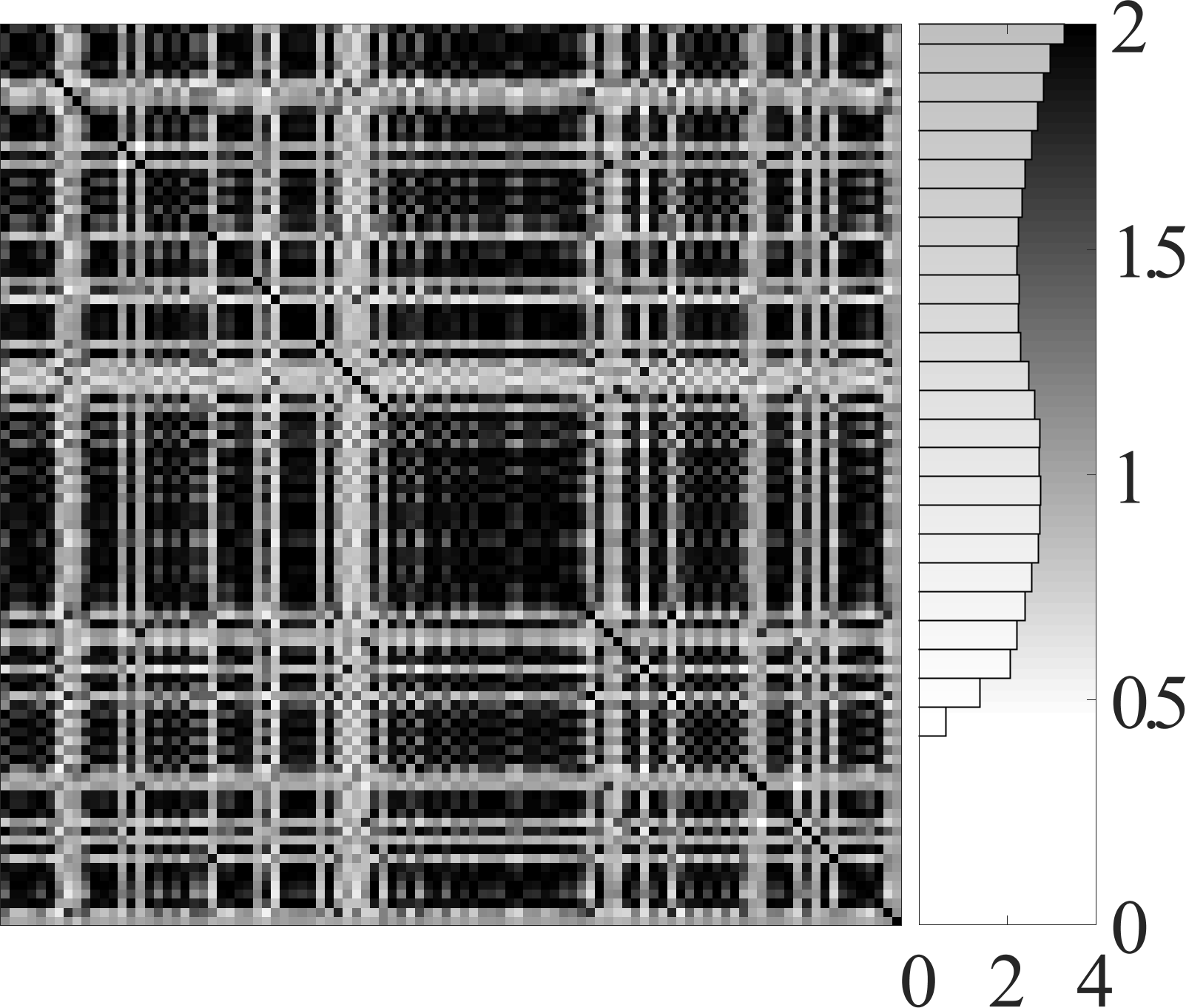}
        \put(25,86){\footnotesize $\eta = 1000$}
        \put(0,86){(f)}
    \end{overpic}
    \caption{
    \label{fig:subgraph_constraint_example_1}
    Subgraph constraint with two subpopulations for Eqs.~\eqref{eq:constrainedAdaptiveKM} resulting in an incoherence-coherence transition.
    Bifurcation diagrams in panels (a) to (c) are quasi-continued from right to left. Initial conditions are described in App.~\ref{app:initial_conditions}. Parameters are $N=2Q$ with $Q=50, \sigma = 0.7068, a=1,b=1, \beta=0, \alpha = 0, \epsilon = 0.1$ and simulation time of $T = 1000$ units. For a detailed description on the measurements and numerical procedures, please refer to Sec.~\ref{sec:SubgraphConstr} and Sec.~\ref{sec:subgraph_incoherence_coherence}.
}

\end{figure}

\begin{figure}
    \begin{overpic}[width=\columnwidth]{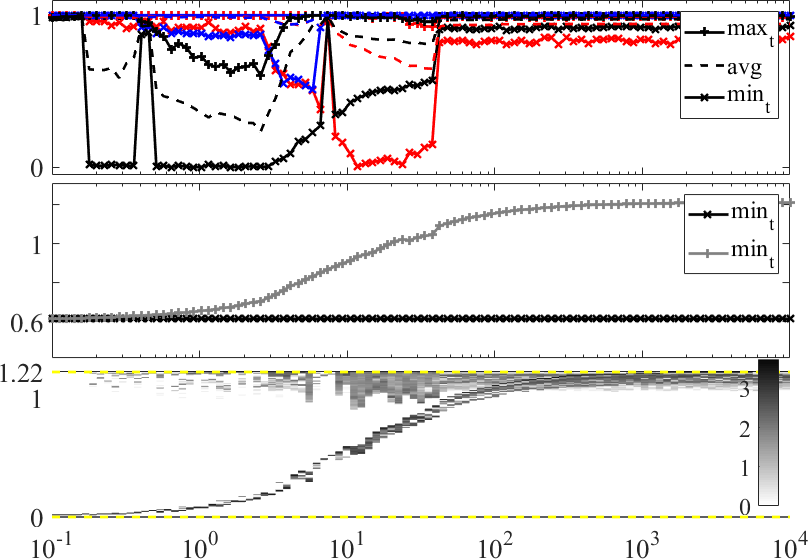}
    \put(50,-1){$\eta$}
    \put(64,56){\footnotesize Partial Sync}
    \put(42,51.5){\footnotesize Chimera}
    \put(15,50.5){\footnotesize 2 pop. clusters}
    \put(-2,48){\footnotesize \rotatebox{90}{${\color{red}{|Z_{1}|}},{\color{blue}{|Z_2|}}, |Z| $}}
    \put(-2,32){\footnotesize \rotatebox{90}{$\langle w_{kj}\rangle$}}
    \put(-2,8){\footnotesize \rotatebox{90}{$w_{kj}(T)$}}
    \setlength{\fboxsep}{1pt}
    \put(86,5){ \rotatebox{90}{ \colorbox{white}{\fbox{  \textcolor{black}{\tiny $\log_{10}(\#)$}} }}}
    \put(7,64){(a)}
    \put(7,43){(b)}
    \put(7,19){(c)}
    \end{overpic}
    \vspace{1em}
    \\
    \begin{overpic}[width=0.32\columnwidth]{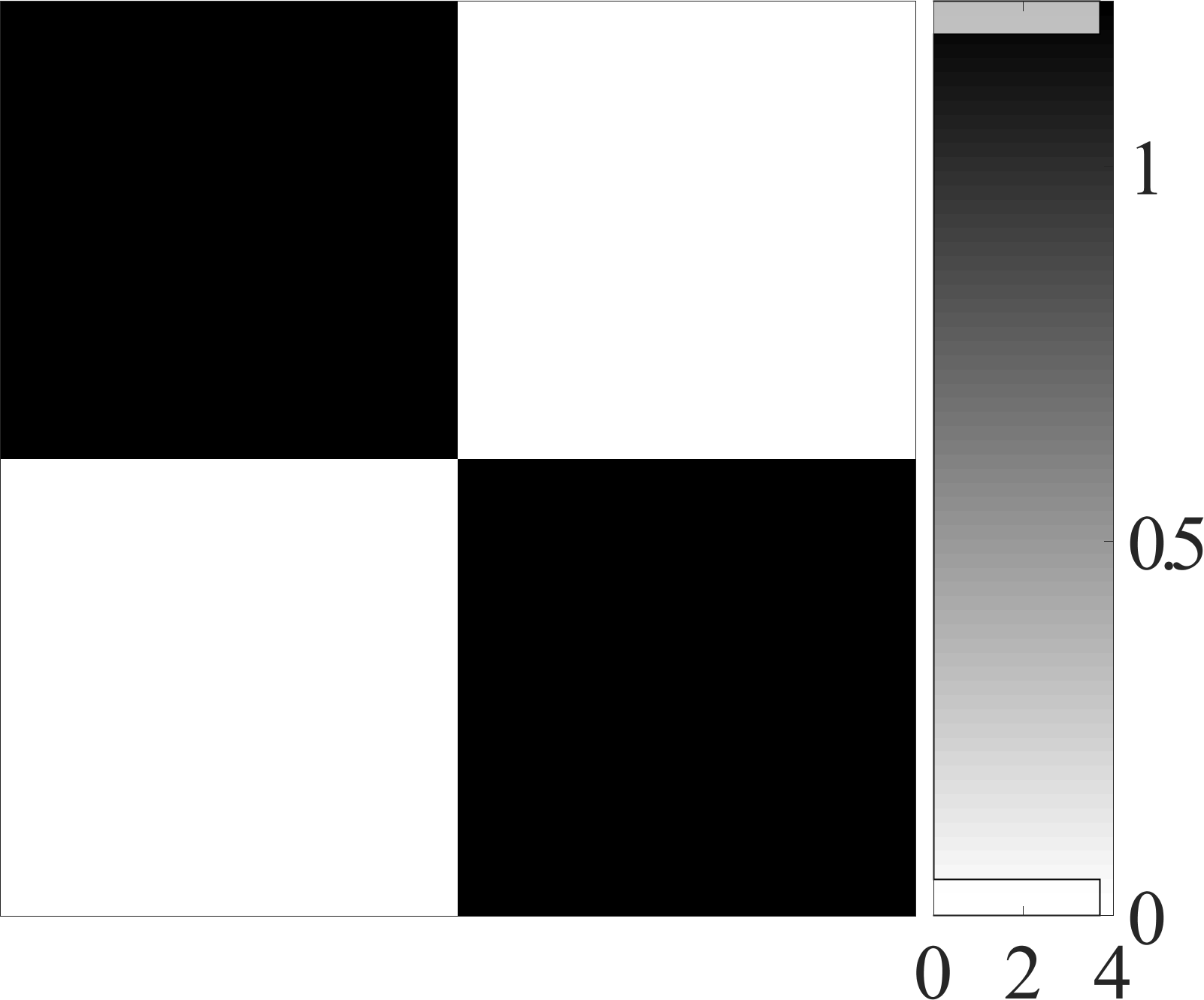}
        \put(25,86){\footnotesize $\eta = 0.64$}
        \put(0,86){(d)}
        \put(73,-6){\tiny \rotatebox{0}{$\log_{10}$}(\#)}
    \end{overpic}
    \begin{overpic}[width=0.32\columnwidth]{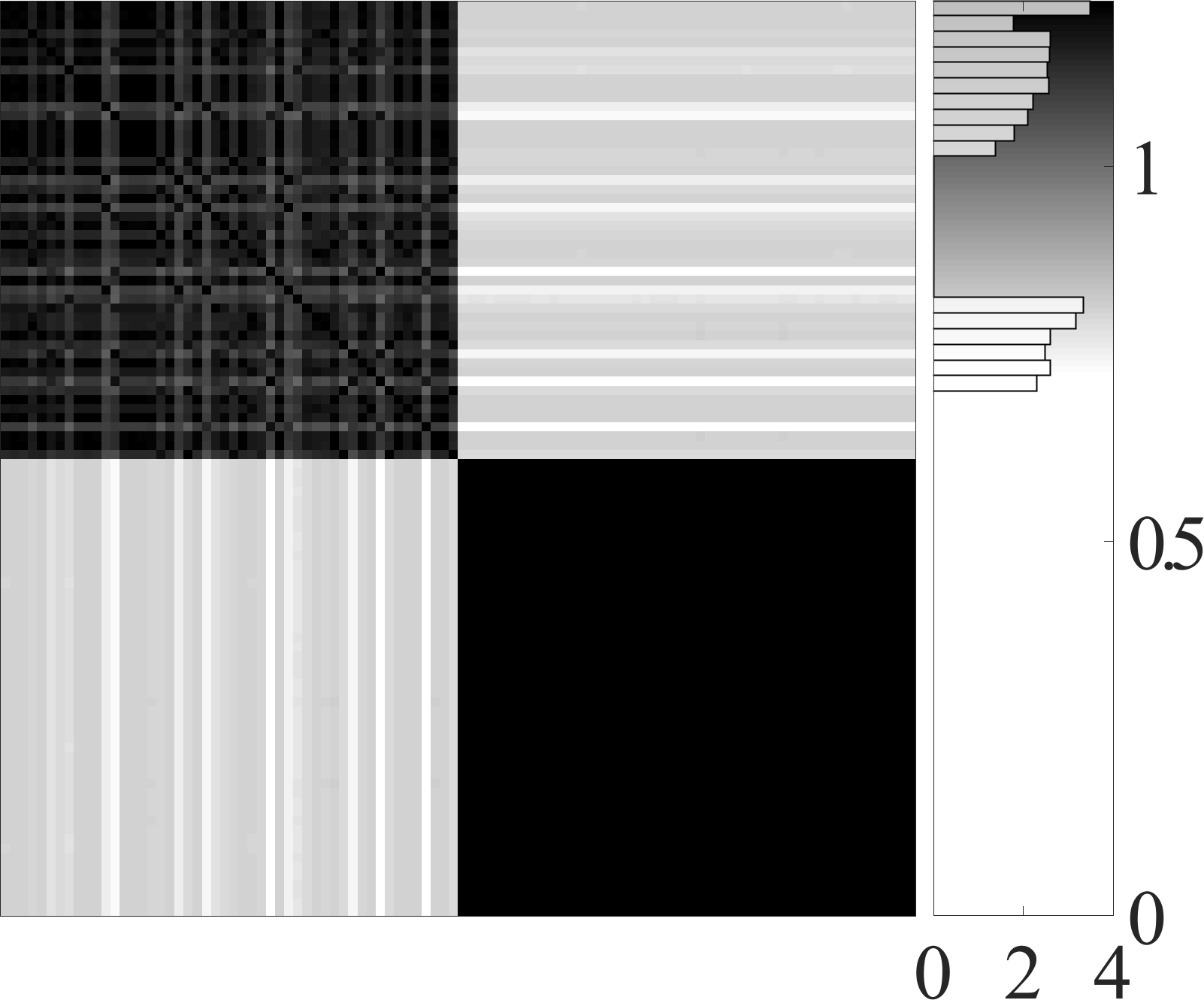}
        \put(22,86){\footnotesize $\eta = 21$}
        \put(0,86){(e)}
    \end{overpic}
    \begin{overpic}[width=0.32\columnwidth]{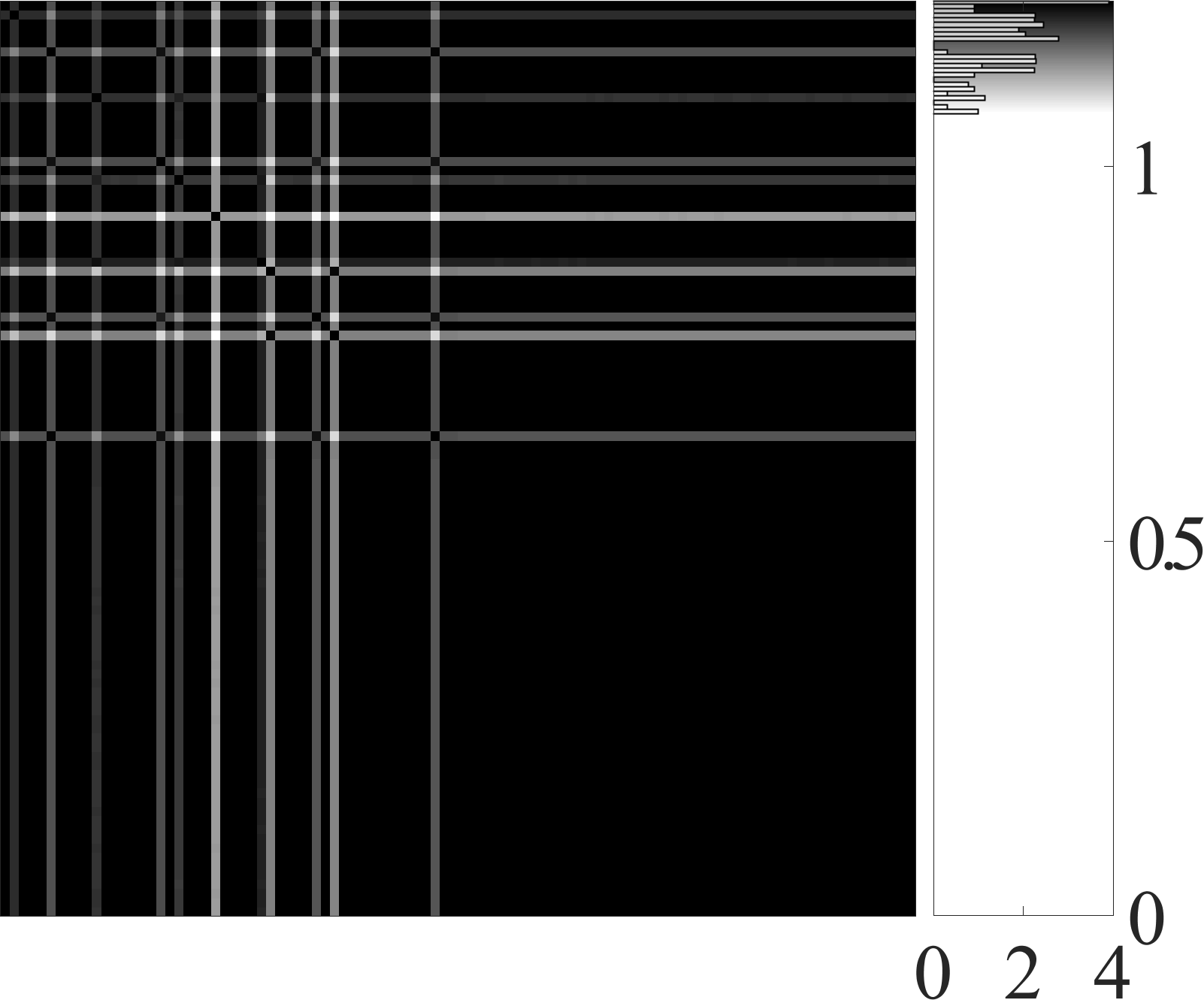}
        \put(24,86){\footnotesize $\eta = 10^4$}
        \put(0,86){(f)}
    \end{overpic}

    \caption{
    \label{fig:subgraph_constraint_example_2}
    Subgraph constraint with two subpopulations for Eqs.~\eqref{eq:constrainedAdaptiveKM} resulting in transitions between 2-cluster, alternating chimera, chimera and partially synchronized states (see Sec.~\ref{sec:subgraph_adaptive_chimera}).
    Parameters are $Q=50, \sigma = 0.001, a=0.22,b=1, \beta=0, \alpha = \pi/2-0.1, \epsilon = 0.1, \eta = 0.1$ and a simulation time of $T = 2000$.
    Identical initial conditions were used for all values of $\eta$ (see App.~\ref{app:initial_conditions}). For details and numerical procedures, see Sec.~\ref{sec:SubgraphConstr}
    ).
    }
\end{figure}
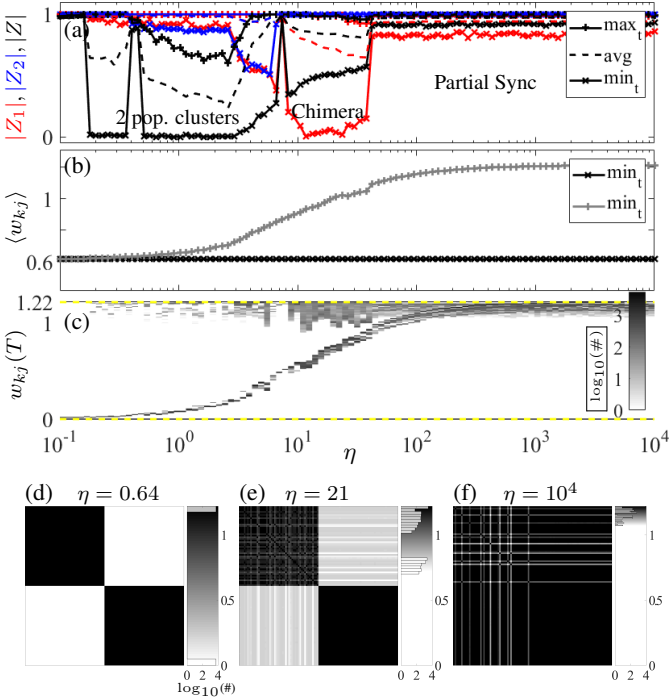

We  now intend to study the dynamic behavior of \eqref{eq:constrainedAdaptiveKM} for the two examples detailed further below, with results illustrated in Figs.~\ref{fig:subgraph_constraint_example_1} and \ref{fig:subgraph_constraint_example_2}. To do this, we shall measure various observable quantities. First, we measure the magnitude of the order parameters to assess the synchronization level for each (sub-)population,
\begin{align}
|Z_1(t)| &:= \frac{1}{Q}\Bigg|\sum_{j=1}^Q e^{i \phi_j(t)}\Bigg|, & |Z_2(t)| &:= \frac{1}{Q}\Bigg|\sum_{j=Q+1}^N e^{i \phi_j(t)}\Bigg|,
\end{align}
as well as the global order parameter $|Z(t)| := \frac{1}{2}|(Z_1(t)+Z_2(t))|$, and we report minima, maxima and averages sampled after a transient time over the interval $[0.5T,T]$ (panels~(a)).
Second, we report minima and maxima of the weights averaged over all edges, $\langle w_{kl}(t)\rangle_{(k,l)} := \sum_{k,l}w_{kl}(t)$, sampled again over the interval~$[0.5T,T]$ (panels~(b)).
Third, we carry out histogram counts for the edge weights~$w_{kl}(T)$ in dependence of the constraint parameter~$\eta$, which we represent as density plots (henceforth referred to as `histograms') while varying values of~$\eta$ as parameter (panels(c)).
Finally, we illustrate weight matrices $w_{kl}(T)$ including bar chart histograms for selected values of $\eta$ (Panels~(d) to~(f)).
Numerical results are obtained using Matlab using initial conditions as described in App.~\ref{app:initial_conditions}.

\paragraph{Incoherence--Coherence transition.\label{sec:subgraph_incoherence_coherence}}

To observe the dynamics studied in Ref.~\onlinecite{cestnik2025continuum},  we choose $\alpha=0, b = 1, \beta =0$ and small~$\epsilon = 0.1$, and in particular, to observe states of partial synchronization,  we choose $\sigma = 0.7068$ and $ a = 1 $.
Bifurcation diagrams in Fig.~\ref{fig:subgraph_constraint_example_1}(a–c) were obtained by numerically integrating Eqs.~\eqref{eq:constrainedAdaptiveKM}) while decreasing $\eta$ from right to left using quasi-continuation and the initial condition specified in App.~\ref{app:initial_conditions}.
We discuss the dynamics observed for different constraint regimes:

{(i) \bf Unconstrained adaptation (partial coherence).}
For (effectively) unconstrained adaptation ($\eta=10^4$), the complete network is globally uniformly adaptive. For the chosen parameters, we then observe order parameters consistent with partial synchronization (coherence), see Fig.~\ref{fig:subgraph_constraint_example_1}(a).
While edge weights fluctuate due to the adaptive co-evolutionary nature of the system (Fig.~\ref{fig:subgraph_constraint_example_1}(b)), they tend to accumulate near $w_{kl}\approx 1,2$ (Fig.~\ref{fig:subgraph_constraint_example_1} panels (c) and (f)).
This can be explained as follows.
Since most phases for partial synchrony are approximately phase synchronized, $\cos(\phi_j-\phi_k)\approx 1$, most edge weights assemble near $w_{kj}\approx a+b =2 $. Drifting oscillators, on the other hand, result in $\cos(\phi_j-\phi_k)\approx 0$, with fluctuation amplitudes scale with the value of the (small but fixed) parameter~$\epsilon$, implying that $w_{kj}\approx b = 1$.

{(ii) \bf Enforced constraint (incoherence).}
For the enforced constraint ($\eta = 0$), we observe that the minimum of the order parameter~$Z(t)$ is close to 0, consistent with incoherent oscillations where the order parameter comes arbitrarily close to zero in the asymptotic time limit.
The constraint enforces the decoupling of the two subpopulations: edge weights connecting the subpopulations tend to zero as $\eta\to 0$; conversely, edge weights connecting vertices within each subpopulation remain non-zero and adaptive, i.e., they co-evolve with the phases associated with these edges. Such a scenario is also seen if $\eta>0$ is small, e.g., see Fig.~\ref{fig:subgraph_constraint_example_1}(d) for $\eta=0.1$.
While for small $\eta>0$, a few oscillator remain locked resulting in a minor accumulation of weights near $w\approx a + b = 2$, the majority of oscillators undergo incoherent oscillations so that the associated phase differences average to zero, implying that associated weights accumulate near $w\approx a = 1$ (Fig.~\ref{fig:subgraph_constraint_example_1}(c)).

{(iii) \bf Intermediate $\eta$-values (incoherence-coherence transition).}
The subgraph structure becomes increasingly prominent in the weights $w_{kj}$ as~$\eta$ decreases, see Fig.~\ref{fig:subgraph_constraint_example_1}, panels (d) to (f). On one hand, enforcing the two-subpopulation subgraph constraint by decreasing $\eta$ effectively halves the coupling strength, leading to the loss of coherent oscillations;
on the other hand, increasing~$\eta$ weakens the  constraint so that an increasing number of edges may co-evolve with their associated phases. Thus, it becomes possible for two oscillators~$k$ and~$j$ to lock, and given that $\beta=0$, associated edge weights are up-regulated. In effect, the average coupling strength of the network is icnreased in this way so that more and more oscillators are recruited into a locked / partially synchronized state with increasing~$\eta$.
The subgraph constraint effectively controls the level of ``accumulated'' coupling.

At a specific value of $\eta$ the minimum value (in time) of the order parameter~$|Z|$ detaches from zero, clearly indicating a transition from incoherent oscillations to partial synchrony.
Note that edge weights consistent with incoherence ($w=0,1$) and partially synchronized states ($w=1,2$) are prominent features of the histograms in Fig.~\ref{fig:subgraph_constraint_example_1}(c) for a large range of $\eta$~values; 
however, weights also accumulate near values in between as they gradually assemble in new quasi-stationary configurations before and after this incoherence-coherence transition.

Simulations using a number of different initial conditions at the beginning of the quasi-continuation indicate that the incoherence-coherence transition does not occur at a unique critical value of~$\eta$. In fact, it is possible to continue an incoherent solution from left to right throughout the range of $\eta$~values shown in the bifurcation diagram. This observation can be attributed to the high multistability observed in the adaptive Kuramoto model (see also Ref.~\onlinecite{cestnik2025continuum}). Consequently, one expects that it is possible that partial synchrony may nucleate for a whole range of $\eta$~values. (Indeed, incoherence-coherence transitions might therefore even be possible multiple times at distinct $\eta$-values.)
Regardless, the structure of the histogrammatic bifurcation diagram for the edge weights~$w$ (Fig.~\ref{fig:subgraph_constraint_example_1}(c)) remains similar; the actual distributions of edge weights~$w_{kj}$ (away from the values $0,1$, and $2$) depends on the various phase configurations the system settles on, given a specific initial condition for the quasi-continuation in~$\eta$. The question regarding the details of this incoherence-coherence transition goes beyond the scope of the present study, and we leave it for a future investigation.

\paragraph{Adaptive chimera states.\label{sec:subgraph_adaptive_chimera}}
To illustrate the effect of the subgraph constraint with two subpopulations, we also investigated parameter values  consistent with observations of chimera states in networks with a (stationary) two subpopulation structure.
(That is, we choose $b=1,a=0.22$ and $\alpha=\pi/2-0.1$, $\beta=0$, so that the amplitude in the coupling modulation, $a$, exceeds $\pi/2 - \alpha$ by a factor of two, see Eq.~[17] in Ref.~\onlinecite{Abrams2008}.)
Bifurcation diagrams in Fig.~\ref{fig:subgraph_constraint_example_2}(a--c) were obtained by numerically integrating Eqs.~\eqref{eq:constrainedAdaptiveKM} using identical initial conditions while varying $\eta$ (see App.~\ref{app:initial_conditions}).
Enforcing the subgraph constraint with $\eta= 0$, the two subpopulations are effectively decoupled (compare with Fig.~\ref{fig:subgraph_constraint_example_2}(d) for $\eta=0.64$), since the constraint with $a_{kj}=0$ quenches edge weights connecting the two subpopulations to $w_{kj}=0$. In contrast, coupling weights within each subpopulation remain adaptive and adhere to values near $w=a+b\cos{\beta} = 1.22$, allowing oscillators within each subpopulation to achieve near-perfect synchrony.
Weights consistent with incoherent oscillations appear absent, possibly because $\alpha\neq 0$ destabilizes incoherent solutions~\cite{sakaguchi1986soluble,Bick2018c}.
In the limit of no constraint, $\eta \to \infty$, the subgraph structure disappears, and the network is effectively globally uniformly adaptive. This enables partial synchrony across the entire network, with disorder in the frequencies or initial  phases persisting in both the phases and coupling weights (compare with Fig.~\ref{fig:subgraph_constraint_example_2}(f) for $\eta=10^4$).
Notably,  edge weights accumulating near $ w \approx 1.22 $  enables some oscillators to phase-lock throughout the entire range of~$\eta$ as shown in Fig.~\ref{fig:subgraph_constraint_example_2}.

For intermediate values of $\eta$, non-trivial dynamics emerge due to the soft subgraph constraint:

 (i) {\bf 2-population clusters / alternating chimera.} 
 For $ \eta\in[0.1,2]$, we observe two frequency-clusters drifting apart. At small~$\eta$, both clusters appear fully coherent. For $\eta\approx 0.2$, coherence in one cluster breaks into partial synchrony while the other stays coherent. 
 These roles alternate periodically, forming a `traveling wave'~\cite{hong2016phase,Hong2021a} or alternating chimeras~\cite{Bick2017c,Zhang2020}. 
 As~$\eta$ increases, both amplitude and frequency of the oscillating order parameters~$|Z_{1,2}|$ grow.  This behavior also occurs for identical oscillator frequencies.
 Over a range of~$\eta$ values, this state is bistable with a state where both closes are locked nearly in-phase (Whether such states persist across the full $\eta$ range remains an open question.)
 
 (ii) {\bf Adaptive Chimera.} For $ \eta\in[7,30]$, one cluster locks into perfect coherence, while the other exhibits periodic oscillations in the order parameter, corresponding to an adaptive (breathing) chimera-like state.
 
 (iii) {\bf Partial Synchrony}. At around $\eta=40$, the system abruptly transitions to a state resembling partial synchrony, where order parameter of the cluster previously oscillating now appears (quasi-)stationary.

Finally, we stress that these states arise due to the varying strength of the subgraph constraint. As $\eta$ decreases, the edge weight distribution $w_{kl}$ bifurcates into two peaks: one near $w=a+b\cos{\beta}=1.22$, and another gradually approaching 0 in a sigmoidal fashion (see Fig.~\ref{fig:subgraph_constraint_example_2} (b) and (c)). This transition from a unimodal to bimodal distribution facilitates the emergence of chimera and alternating two-cluster states.

\subsection{Constant input/output constraint (degree constraint)}

\noindent
Finally, we consider a constraint that allows connectivity to vary while keeping the (weighted) in-degree of a node constant.
This is given by a row sum constraint~\eqref{eq:RowSumConstr},
\begin{align*}
\Ct &= \set{\w=(w_{kj})}{\textstyle\sum_{j\neq k}w_{kj}=c_k}.
\end{align*}
For example, the row sums~$c_k$ could be seen as sampled from a (weighted) in-degree distribution $\mathcal{D}$.
To write the constraint as in Sec.~\ref{sec:FormalEquations}, the constant row sum offset is given by
\begin{align}
\ws = 
\begin{pmatrix}
0 & \frac{c_1}{N-1} & \cdots & \frac{c_1}{N-1} \\
\frac{c_2}{N-1} & 0 & \cdots & \frac{c_2}{N-1} \\
\vdots & \vdots & \ddots & \vdots \\
\frac{c_N}{N-1} & \frac{c_N}{N-1} & \cdots & 0
\end{pmatrix}.
\end{align}
The $N$~directions transverse to~$\Ct$ along~$\U$ are determined by matrices with ones in a given row (except the diagonal). 
With $\bar w_k = \sum_{j\neq k} w_{kj}$ the projections $\w_\V = \Pi_\V(\w-\ws), \w_\U = \Pi_\U(\w-\ws)$ are
\begin{subequations}\label{eq:ConstrProj1}
\begin{align}
\begin{split}
\w_\U &=
 \begin{pmatrix}
0 & \bar w_1-c_1 & \cdots & \bar w_1-c_1 \\
\bar w_2-c_2 & 0 & \cdots & \bar w_2-c_2 \\
\vdots & \vdots & \ddots & \vdots \\
\bar w_N-c_N & \bar w_N-c_N & \cdots & 0
\end{pmatrix},
\end{split}
\\
\begin{split}
\w_\V
&= \w-\ws - \w_\U,
\end{split}
\end{align}
\end{subequations}
so that 
$\w_\Ct = \ws + \w_\V\in \Ct$ and $\w = \w_\Ct + \w_\U$.

We sidestep the complication of computing the explicit projections of the dynamical equations~\eqref{eq:AdaptConstr}. 
Instead, we use a similar approach as in~\eqref{eq:constrainedAdaptiveKM} and again consider the adaptive Kuramoto model. 
Write $\w_\Ct = (w^\Ct_{kj})$.
One way to guarantee that the system satisfies a given degree distribution $c_k\sim\mathcal{D}$ is to consider
\begin{subequations}\label{eq:constrained_degree_AdaptiveKM}
    \begin{align}\label{eq:constrained_degree_AdaptiveKM_phi}
    \dot{\phi}_k &= \omega_k + \frac{1}{N}\sum_{j=1}^N w_{kj} \sin(\phi_j-\phi_k+\alpha), \,\\
    \begin{split} \label{eq:constrained_degree_AdaptiveKM_w}
        \dot{w}_{kj} &= A(\phi_k,\phi_j,w_{kj}) - \eta^{-1}(w_{kj}-w^\Ct_{kj}).
    \end{split}
    \end{align}
\end{subequations}
Thus, $\eta \to 0 $ guarantees that the weights exactly obey the (given) realization of the weighted degrees, $c_k\sim\mathcal{D}$, and for $\eta \to \infty$ the system is again uniformly globally adaptive.

\begin{figure}
 \begin{overpic}[width=0.45\columnwidth]{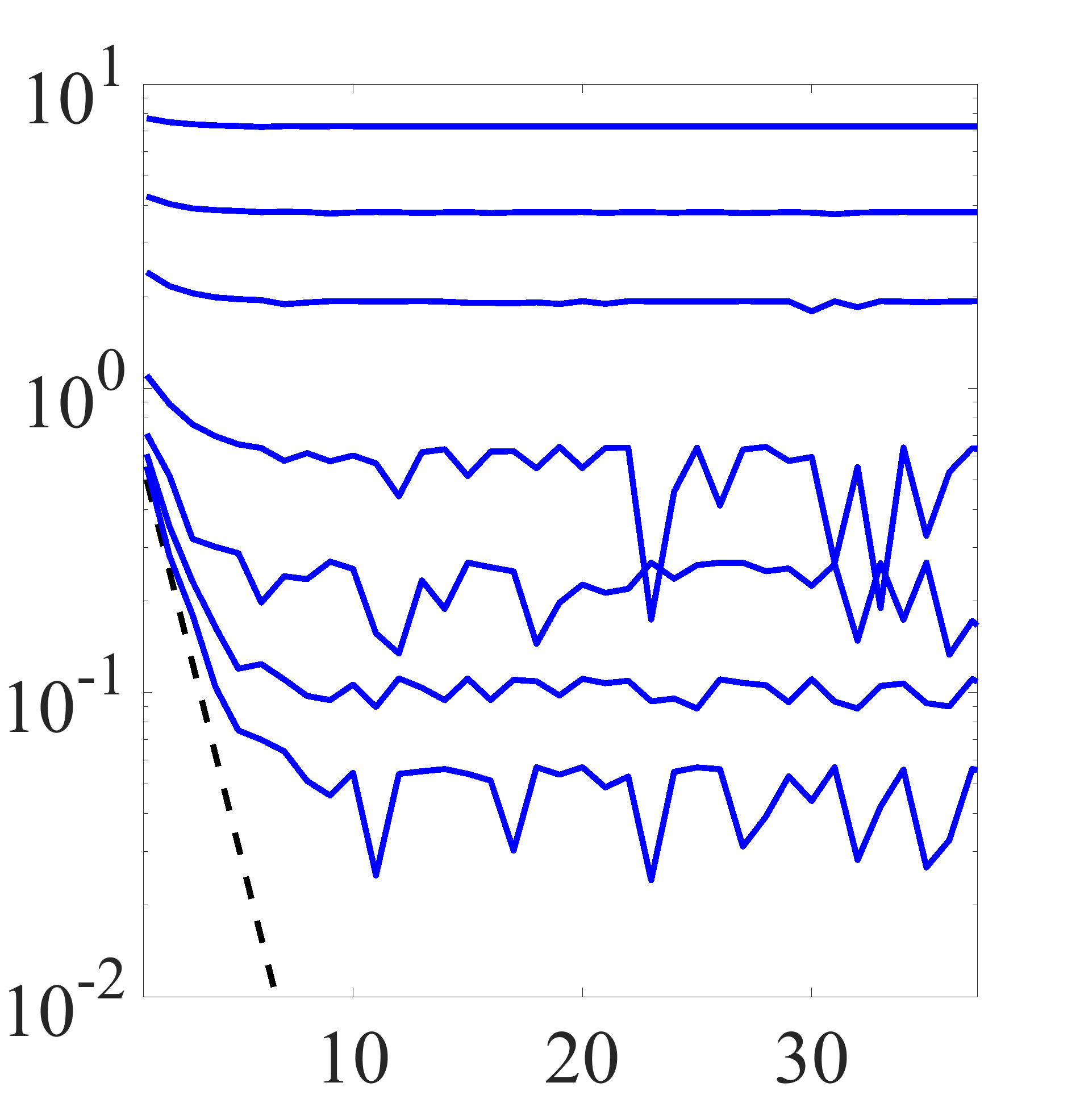}
  \put(-8,30){\rotatebox{90}{Degree $w_k$}}  
   \put(40,-5){Index $k$}  
   \put(34,90){\tiny $\eta=1$}
   \put(34,83){\tiny $\eta=0.5$}
   \put(34,75){\tiny $\eta=0.25$}
   \put(34,64){\tiny $\eta=0.1$}
   \put(34,51){\tiny $\eta=0.5$}
   \put(34,40){\tiny $\eta=0.025$}
   \put(34,32){\tiny $\eta=0.01$}
   \put(12.5,40){\rotatebox{-77}{\tiny Constraint}}
   \put(40,95){$\gamma=0.5$}
 \end{overpic}
 \hspace{1em}
 \begin{overpic}[width=0.45\columnwidth]{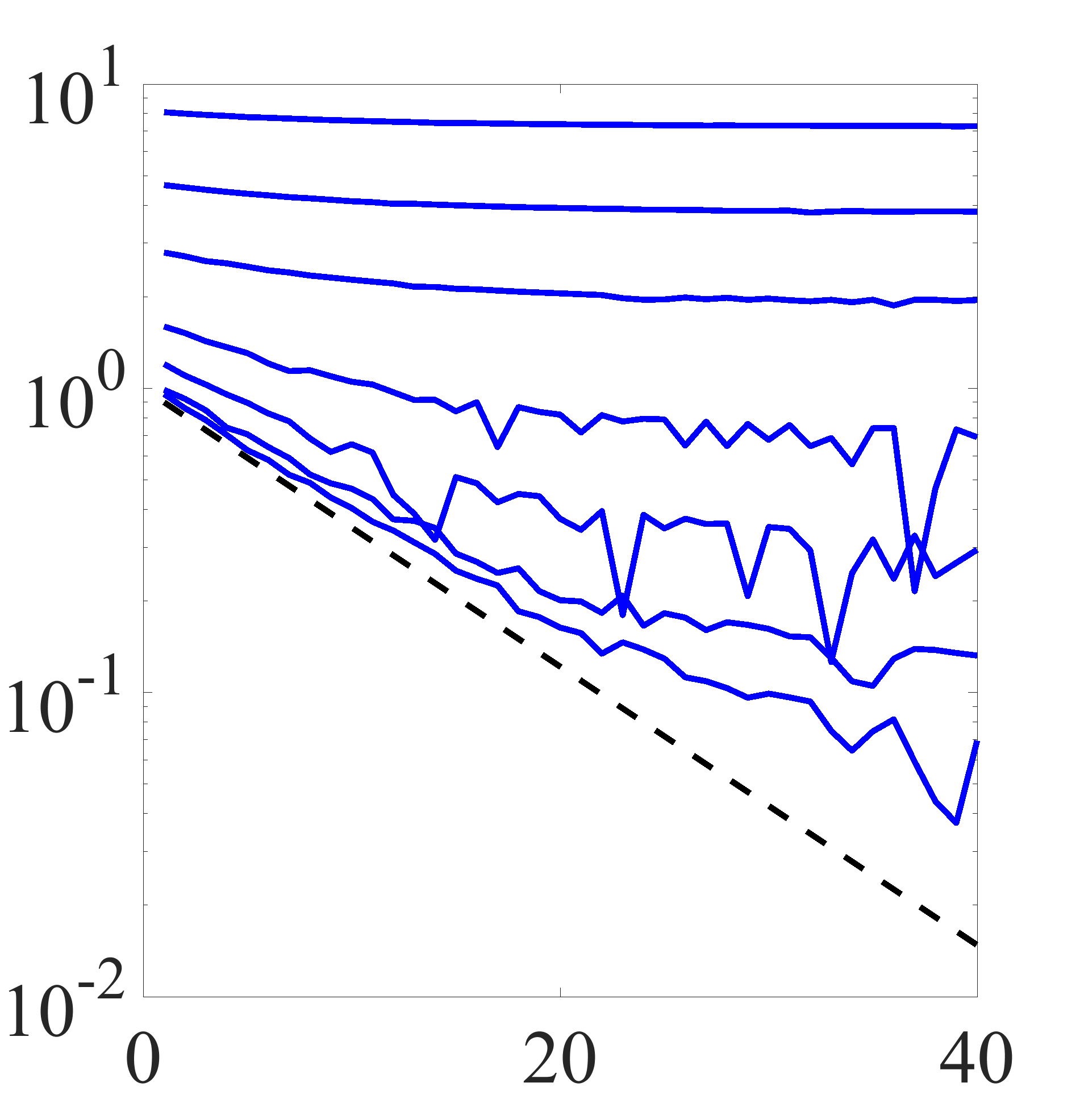}
  \put(-8,30){\rotatebox{90}{Degree $w_k$}}  
   \put(40,-5){Index $k$}  
   \put(64,89){\tiny $\eta=1$}
   \put(64,83){\tiny $\eta=0.5$}
   \put(64,74){\tiny $\eta=0.25$}
   \put(64,63){\tiny $\eta=0.1$}
   \put(64,54){\tiny $\eta=0.5$}
   \put(64,43){\tiny $\eta=0.025$}
   \put(64,33){\tiny $\eta=0.01$}
   \put(27,50){\rotatebox{-35}{\tiny Constraint}}
   \put(40,95){$\gamma=0.9$}
 \end{overpic}

 \caption{\label{fig:degree_constraint}
The homeostasis time scale~$\eta$ determines the convergence behavior to a weighted in-degree constraint~$\Ct$ determined by $c_k=K \gamma^k$ for each node $k=1, \dotsc, N$.
The panels show the weight distribution for different values of the exponent~$\gamma$ and fixed $K=1$  after $T=4000$ time units for an adaptive Kuramoto network~\eqref{eq:constrained_degree_AdaptiveKM} of $N=40$ oscillators with parameters $\epsilon=0.1$, $\alpha=\beta=0$, $a=b=1$, $\sigma=0.01$.
}
\end{figure}
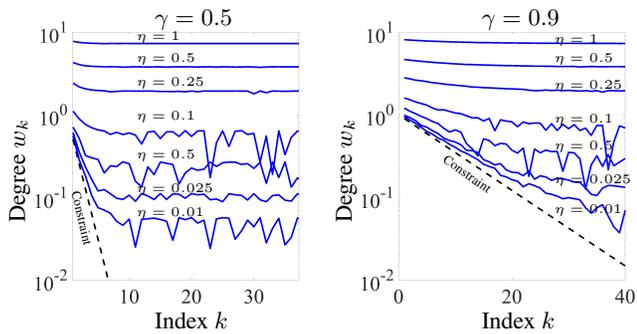

To illustrate the effect of this constraint, we match the row sums with a power law distribution, i.e., $c_k=K \gamma^k$ with $k= 1,\ldots, N$. 
Thus, coupling weights consistent with the constraint approximate the structure of a (weighted) scale-free network. 
Results for such constraints are shown for two different values of~$\gamma$ in Fig.~\ref{fig:degree_constraint}.
Simulations are carried out for $\gamma=0.5$ and 0.9 while varying~$\eta$. As expected, the constraint is matched better as the constraint hardens towards the asymptotic $\eta\to 0$. Moreover, one observes that the convergence to the constraint worsens for increasing row index~$k$, presumably due to the time scales associated with~$c_k$ depending on~$k$.
Similarly, better convergence to the constraint is achieved by either chosing a larger value of $\gamma$ (compare $\gamma=0.9$ with $\gamma=0.5$), or by increasing the scaling factor~$K$ of the degree distribution.

\section{Discussion}
\label{sec:Discussion}

\noindent
We have introduced a versatile framework for constrained adaptation (Sec.~\ref{sec:constrained_adaptive_network_dynamics}),
which extends existing approaches for adaptive/co-evolutionary network dynamics~\cite{gross2008adaptive,berner2023adaptive,sawicki2023perspectives}. This framework is broadly applicable to systems described by Eqs.~\eqref{eq:AdaptiveNetwork} and it accommodates a wide range of adaptive/co-evolutionary network models. 
To illustrate its flexibility, we provided a non-exhaustive list of example constraints (Sec.~\ref{sec:constraint_examples}), covering constraints at the network level, node level, and within a given subgraph. 
These constraints can either be enforced via constants or be allowed to evolve according to their own adaptive dynamics. 
For example, the adaptive dynamics that arise from global adaptive coupling, as defined in Eq.~\eqref{eq:constraint_global_adaptive}, and from adaptive input weights, as in Eq.~\eqref{eq:constraint_adaptive_input}, can be studied in their own right. These have been studied by Duchet {\it et al.}~\cite{Duchet2023} and by  Cestnik and Martens~\cite{cestnik2025continuum}, respectively. 
We demonstrated that symmetries inherent to the adaptation dynamics can give rise to \emph{intrinsic constraints}, as discussed in (Sec.~\ref{sec:KMN2_forcedconstraints}). 
These constraints arise naturally when the system evolves toward an attractive subspace determined by the symmetries of the adaptation process. 
In contrast, \emph{forced constraints} (homeostasis) reduce the system's dimensionality  by confining the dynamics to a predefined constraint manifold, regardless of any underlying symmetry. 
This distinction between intrinsic and forced constraints is central to understanding how different adaptive processes shape the system's long-term behavior (Sec.~\ref{sec:KMN2_forcedconstraints}).

In general settings, particularly in large or heterogeneous networks, one may not expect to find perfectly intrinsic constraints to arise. 
Nevertheless, it may still be useful to restrict the dynamics to an optimal constraint manifold, effectively reducing the system's dimensionality. 
An intuition that certain directions in the adaptation space may be less important, or knowledge of approximate symmetries, can inform a suitable choice for a constraint manifold. 
If this constraint manifold can be suitably parameterized, one could envision an \emph{optimization framework} that minimizes the dynamics along $\U$ with respect to a suitable norm. 
Even when the adaptive dynamics along this direction do not vanish, it can be further suppressed/quenched by activating an appropriate constrained dynamics. This approach suggests a hybrid method---partly analytical, partly data-informed---for simplifying and controlling high-dimensional adaptive dynamics.

This perspective connects naturally to the observations made regarding the forced constraint for the adaptive Kuramoto model with $N=2$ oscillators. 
In the limit $\eta \to 0$, adaptation is strictly confined to a lower-dimensional (critical) constraint manifold. For small $\eta>0$, the dynamics deviate slightly from this manifold, yet still effectively evolve within a space compatible with its dimensionality. 
As $\eta$ increases further, bifurcations emerge, enabling higher-dimensional dynamics---analogous to scenarios where the adaptivity~$a$ is gradually increased~\cite{juttner2023complex,augustsson2024co,cestnik2025continuum}. This behavior is well illustrated in the case of two adaptive Kuramoto oscillators (Eqs.~\eqref{eq:N2adaptiveKM}), which undergo a transition from periodic to chaotic dynamics via a period-doubling cascade as $\eta$~increases. 
An explicit analysis for $N\geq 3$ oscillators---for example how such transitions arise---would be interesting for future work, but the dimensionality gives rise to many possible constraints.

For larger networks, we demonstrated how subgraph and degree constraints influence the system's dynamics. Subgraph constraints imprint structural features onto the network, modifying the overall coupling strength. As the constraint parameter varies, these subgraph constraints can drive transitions from incoherence to coherence, or induce more intricate transitions involving two-cluster states, chimera-like patterns and partial synchrony, depending on parameter choices. 
In these more intricate regimes, the effective dimensionality of the dynamics changes in a nontrivial way. 
At intermediate values of~$\eta$, the system exhibits two-cluster or chimera-like states, composed of two interacting macroscopic oscillators~\cite{Martens2009}. At larger~$\eta$, partial synchrony emerges, resembling the behavior of a single macroscopic oscillator. However, this macroscopic perspective---based on reductions such as the Ott--Antonsen ansatz~\cite{Bick2018c}---neglects the heterogeneity in the evolving coupling weights, which breaks the symmetries required for such reductions.
Lastly, in the case of degree constraints, the dynamics are characterized by the emergence of multiple time scales.

In addition to the homeostasis time-scale~$\eta$ considered here, adaptive network dynamical systems often involve additional time scales. 
A concrete example is the time-scale separation~$\epsilon$ in the adaptive Kuramoto equations~\eqref{eq:full_system}, which we assumed to be fixed at a finite value throughout our analysis.
However, it is also natural to take simultaneous limits of $\eps\to 0$ and $\eta\to 0$ and identify relevant scaling relationships between these small parameters; see~Ref.~\onlinecite{Kuehn2020a}.
In this simultaneous limit, the adaptation of the network connection is small while the homeostasis that enforces the constraint is fast.
Finally, there is a natural limit of $N\to\infty$ that describes a continuum limit\cite{Medvedev2013a} of infinitely many adaptive oscillators\cite{Duchet2023,gkogkas2023continuum}.

Although our framework is currently formulated for systems described by Eqs.~\eqref{eq:AdaptiveNetwork} and we limited our study to the constraints listed in Sec.~\ref{sec:constraint_examples}, the framework is readily extended. 
Potential generalizations include
(i)~non-pairwise interactions, such as those found in higher-order networks~\cite{bick2023higher}; 
(ii)~alternative constraint types, including curved constrait manifolds (with appropriate care near singularities);
(iii)~weighted subgraph constraints, or replacing the baseline $b$ with a weighted graph;
(iv)~heterogeneous nodal dynamics, distinct from the homogeneous nodal dynamics given by Eqn.~\eqref{eq:NodeDyn}, such as metabolic or gene expression networks~\footnote{Noting, however, that embedding networks with dissimilar dynamics, such as metabolic or gene expression networks, into a framework of adaptive networks and constraints poses a significant challenge.}.

Ultimately, an interesting direction involves extending constrained adaptation to continuous graph representations such as graphons or graphops~\cite{Chiba2019,Kuehn2020}. Adaptive dynamics in these formulations have only begun to be researched recently and present certain theoretical challenges~\cite{gkogkas2022mean,gkogkas2023continuum}. 
Beyond the generalizations already discussed, future work could explore real-world applications involving multiple adaptation time scales, as  may occur with synaptic plasticity in neuroscience, machine learning, control, and power grid structures.

\begin{acknowledgments}
\noindent
    We gratefully acknowledge financial support from the Royal Swedish Physiographic Society of Lund.
    We are grateful for helpful comments from reviewers.
\end{acknowledgments}

\section*{Data availability statement}
\noindent
Data sharing is not applicable to this article as no new data were created or analyzed in this study. Code will be made available upon reasonable request.

\appendix

\section{Initial conditions\label{app:initial_conditions}}

\paragraph{Incoherence-Coherence transition.}
Initial conditions for the simulations described in Sec.~\ref{sec:subgraph_incoherence_coherence} (Fig.~\ref{fig:subgraph_constraint_example_1}) are $\phi_k(0)\sim\pi/2 + 0.5 \cdot \mathcal{N}(0)$; and 
\begin{align*}
\w(0) = 
        \left(
        \begin{array}{cc}
         \boldsymbol{\Db_1} & \boldsymbol{\Bb}\\
         \boldsymbol{\Bb}^\tr & \boldsymbol{\Db_2}
        \end{array}
        \right) \in \R^{N\times N}, 
\end{align*}
where 
$\Db_1\in\R^{N_1\times N_1} \sim  \mathcal{N}(1,0.05)$,
$\Db_2\in\R^{N_2\times N_2} \sim  \mathcal{N}(2,0.05)$, 
$\Bb \in\R^{N_2\times N_1} \sim  \mathcal{N}(1,0.05)$,
with
$N_1 = \lceil 0.8 \cdot N \rceil$
and 
$N_2 = N - N_1$.

Bifurcation diagrams in Fig.~\ref{fig:subgraph_constraint_example_1} panels (a) to (c) are quasi-continued in $\eta$ from right to left, using this initial condition.

\paragraph{Adaptive Chimeras.}
Initial conditions for the simulations described in Sec.~\ref{sec:subgraph_adaptive_chimera} (Fig.~\ref{fig:subgraph_constraint_example_2}) are 
$\phi_k(0)\sim \mathcal{N}(-0.5,0.05)$ for $ k=1,\ldots, Q$ and  
$\phi_k(0)\sim \mathcal{N}(0,0.05)$ for $k = Q+1,\ldots, N$ where $N=2Q$; and
\begin{align*}
\w(0) = 
        \left(
        \begin{array}{cc}
         \boldsymbol{\Ab} & \boldsymbol{\Bb}\\
         \boldsymbol{\Bb}^\tr & \boldsymbol{\Ab}
        \end{array}
        \right) \in \R^{N\times N}, 
\end{align*}
where 
$\Ab = (a_{kl}) \in \R^{Q\times Q}, a_{kl} \sim \mathcal{N}(1+a,0.1)$;
$\Bb = (b_{kl}) \in \R^{Q\times Q}, b_{kl} \sim \mathcal{N}(1-a,0.1)$.

Bifurcation diagrams in Fig.~\ref{fig:subgraph_constraint_example_2} panels (a) to (c) use this (same) initial condition for every value of $\eta$.

\bibliographystyle{apsrev4-2}

\end{document}